\documentclass{article}

\input{tcilatex}

\begin{document}

\title{{\LARGE Biorthogonal\ Polynomials for Potentials of two\ Variables
and External\ Sources at the Denominator}}
\author{{\Large M.C. BERGERE } \\
Service de Physique th\'{e}orique, CEA-Saclay\\
F-91191 Gif sur Yvette, France\\
email: bergere@spht.saclay.cea.fr}
\date{April 10, 2004}
\maketitle

\begin{abstract}
We construct biorthogonal polynomials for a measure over the complex plane
which consists in the exponential of a potential $-V\left( z,z^{\ast
}\right) \ $and in a set of external sources at the numerator and at the
denominator.\ We use the pseudonorm of these polynomials to calculate the
resolvent integral for correlation functions of traces of powers of complex
matrices (under certain conditions).
\end{abstract}

\bigskip

\bigskip

\bigskip

\section{Introduction}

\qquad In several domains of physics like, for instance, in string theory $%
\left[ 1\right] $ and in quantum chromodynamics $\left[ 2\right] $ we are
interested in the spectral properties of random matrices; most results in
this domain are included in the correlation functions for the invariants of
the matrices like determinants, traces...Although most interesting results
are expected from infinitely large matrices, exact results can be obtained
in appropriate cases for finite matrices using the technique of orthogonal
polynomials. The appropriate cases are when the integrals over the matrix
elements can be transformed into integrals over the eigenvalues of the
matrices; this is the case for hermitian matrices where a unitary
transformation reduces the integrals to the real line for each eigenvalue.
When the matrices are complex, there exists a small class of potentials $%
\left[ 3\right] \ $(which includes the Gaussian potential) where the
integration can be reduced to the complex plane (or a compact domain of the
complex plane) for each eigenvalue. Moreover this is valid only for a
restricted class of correlation functions which is essentially%
\begin{equation}
<\dprod\limits_{i=1}^{p}\left[ Tr\ M^{J_{i}}\right] \ \
\dprod\limits_{i=1}^{q}\left[ Tr\ \left( M^{+}\right) ^{K_{i}}\right] >\ \ \
\ \ \ \ \ \ J_{i}\ \text{and }K_{i}>0  \tag{$1$}
\end{equation}%
in a potential $V\left( M,M^{+}\right) ,\ $which becomes%
\begin{equation}
<\dprod\limits_{i=1}^{p}\left[ \sum_{j}z_{j}^{J_{i}}\right] \ \
\dprod\limits_{i=1}^{q}\left[ \sum_{j}\left( z_{j}^{\ast }\right) ^{K_{i}}%
\right] >  \tag{$2$}
\end{equation}%
in a potential $V\left( z,z^{\ast }\right) .\ $In this context, given a
potential $V\left( z,z^{\ast }\right) $ such that the partition function is
defined as%
\begin{equation}
Z=\int \dprod\limits_{i=1}^{N}d^{2}z_{i}\ \ \dprod\limits_{i<j}\left\vert
z_{i}-z_{j}\right\vert ^{2}\ \ \ e^{-\sum_{i=1}^{N}V\left( z_{i},z_{i}^{\ast
}\right) }  \tag{$3$}
\end{equation}%
we wish to calculate the following integrals%
\begin{equation}
I_{N}=\frac{1}{Z}\int \dprod\limits_{i=1}^{N}d^{2}z_{i}\ \
\dprod\limits_{i<j}\left\vert z_{i}-z_{j}\right\vert ^{2}\ \
\dprod\limits_{j=1}^{N}\left\{ \frac{\dprod\limits_{i=1}^{L_{2}}\left(
z_{j}^{\ast }-\eta _{i}^{\ast }\right) }{\dprod\limits_{i=1}^{M_{2}}\left(
z_{j}^{\ast }-x_{i}^{\ast }\right) }\ \frac{\dprod\limits_{i=1}^{L_{1}}%
\left( z_{j}-\xi _{i}\right) }{\dprod\limits_{i=1}^{M_{1}}\left(
z_{j}-y_{i}\right) }\right\} \ \ e^{-\sum_{i=1}^{N}V\left( z_{i},z_{i}^{\ast
}\right) }\   \tag{$4$}
\end{equation}%
In the special case where $L_{1}=M_{1}$ and $L_{2}=M_{2}\ \ $the application
of\ 

\ $\dprod\limits_{i=1}^{M_{2}}\left( -\frac{\partial }{\partial \eta
_{i}^{\ast }}\right) _{x_{i}^{\ast }=\eta _{i}^{\ast }}\ \
\dprod\limits_{i=1}^{M_{1}}\left( -\frac{\partial }{\partial \xi _{i}}%
\right) _{y_{i}=\xi _{i}}\ \ $on $I_{N}\ $\ defines the resolvent%
\begin{equation}
J_{N}=\frac{1}{Z}\int \dprod\limits_{i=1}^{N}d^{2}z_{i}\ \
\dprod\limits_{i<j}\left\vert z_{i}-z_{j}\right\vert ^{2}\
\dprod\limits_{i=1}^{M_{2}}\left\{ \sum_{j=1}^{N}\frac{1}{z_{j}^{\ast }-\eta
_{i}^{\ast }}\right\} \ \dprod\limits_{i=1}^{M_{1}}\left\{ \sum_{j=1}^{N}%
\frac{1}{z_{j}-\xi _{i}}\right\} \ e^{-\sum_{i=1}^{N}V\left(
z_{i},z_{i}^{\ast }\right) }  \tag{$5$}
\end{equation}%
Clearly enough, the large $\xi _{i},\ \eta _{i}^{\ast }$ expansion of $J_{N}$
is a formal power series the coefficients of which are the correlation
functions $\left( 2\right) .$

\bigskip

In this publication, we calculate $I_{N}$\ and $J_{N};$ the formal power
series of $J_{N}$ will be done somewhere else.\ We now describe the method
which is based on the existence of biorthogonal polynomials for the
potential $V\left( z,z^{\ast }\right) $ in the presence of external sources$%
.\ $

\bigskip

We consider a real potential$\ V\left( z,z^{\ast }\right) $ which admits an
infinite set $\left\{ p_{n}\left( z\right) \right\} $ of orthogonal monic
polynomials%
\begin{equation}
\int d^{2}z\ \ p_{m}^{\ast }\left( z\right) \ \ p_{n}\left( z\right) \ \
e^{-V\left( z,z^{\ast }\right) }=h_{n}\ \delta _{nm}  \tag{$6$}
\end{equation}%
where $p_{m}^{\ast }\left( z\right) \ $is a short notation for $\left[
p_{m}\left( z\right) \right] ^{\ast }.$ More generally, we introduce a
positive Borel measure $\mu \left( z,z^{\ast }\right) $ on the complex plane
and write $\left( 6\right) $ as%
\begin{equation}
\int d\mu \left( z,z^{\ast }\right) \ \ p_{m}^{\ast }\left( z\right) \ \
p_{n}\left( z\right) \ \ =h_{n}\ \delta _{nm}  \tag{$7$}
\end{equation}

We introduce four operations:\bigskip

\qquad \qquad \qquad \qquad \qquad \textit{operation1 :} $d\mu _{1}\left(
z,z^{\ast }\right) =\left( z-\xi \right) \ d\mu \left( z,z^{\ast }\right) $

\qquad \qquad \qquad \qquad \qquad \textit{operation2 : }$d\mu _{2}\left(
z,z^{\ast }\right) =\left( z^{\ast }-\eta ^{\ast }\right) \ d\mu \left(
z,z^{\ast }\right) $

\qquad \qquad \qquad \qquad \qquad \textit{operation3 : }$d\mu _{3}\left(
z,z^{\ast }\right) =\frac{1}{z-y}\ d\mu \left( z,z^{\ast }\right) $

\qquad \qquad \qquad \qquad \qquad \textit{operation4 : }$d\mu _{4}\left(
z,z^{\ast }\right) =\frac{1}{z^{\ast }-x^{\ast }}\ d\mu \left( z,z^{\ast
}\right) $

\bigskip

Although these four measures are not real, each of them admits an infinite
set of monic biorthogonal polynomials. The successive iterations of the four
operations construct an infinite set of monic biorthogonal polynomials for
the measure%
\begin{equation}
d\mu \left( z,z^{\ast };\xi _{i},\eta _{i}^{\ast };y_{i},x_{i}^{\ast
}\right) =\frac{\dprod\limits_{i=1}^{L_{2}}\left( z^{\ast }-\eta _{i}^{\ast
}\right) }{\dprod\limits_{i=1}^{M_{2}}\left( z^{\ast }-x_{i}^{\ast }\right) }%
\ \frac{\dprod\limits_{i=1}^{L_{1}}\left( z-\xi _{i}\right) }{%
\dprod\limits_{i=1}^{M_{1}}\left( z-y_{i}\right) }\ \ d\mu \left( z,z^{\ast
}\right)  \tag{$8$}
\end{equation}%
These monic biorthogonal polynomials are denoted by $q_{n}\left( z;\xi
_{i},\eta _{i}^{\ast };y_{i},x_{i}^{\ast }\right) $ and $q_{n}^{\ast }\left(
z;\eta _{i},\xi _{i}^{\ast };x_{i},y_{i}^{\ast }\right) \ $and satisfy%
\begin{equation}
\int d\mu \left( z,z^{\ast };\xi _{i},\eta _{i}^{\ast };y_{i},x_{i}^{\ast
}\right) \ \ q_{m}^{\ast }\left( z;\eta _{i},\xi _{i}^{\ast
};x_{i},y_{i}^{\ast }\right) \ \ q_{n}\left( z;\xi _{i},\eta _{i}^{\ast
};y_{i},x_{i}^{\ast }\right) =\left\Vert q_{n}\right\Vert ^{2}\ \delta _{nm}
\tag{$9$}
\end{equation}%
Clearly, $q_{n}^{\ast }\left( z;\eta _{i},\xi _{i}^{\ast };x_{i},y_{i}^{\ast
}\right) $ is deduced from $q_{n}\left( z;\xi _{i},\eta _{i}^{\ast
};y_{i},x_{i}^{\ast }\right) \ $by complex conjugation and by the exchange $%
\xi _{i}\Longleftrightarrow \eta _{i},\ y_{i}\Longleftrightarrow x_{i}$
which implies $L_{1}\Longleftrightarrow L_{2},\ M_{1}\Longleftrightarrow
M_{2}.\ $The unicity of the biorthogonal polynomials requires that we stay
away from the surfaces of the $\left( \xi _{i},\eta _{i}^{\ast
};y_{i},x_{i}^{\ast }\right) $ space where$\ \left\Vert q_{n}\right\Vert
=0.\ $

\bigskip

This construction is a generalization of the well-known result by
Christoffel $\ \left[ 4\right] $\ which shows that, given a positive Borel
measure of one variable $d\mu \left( x\right) \ $on the real line and its
infinite set of orthogonal polynomials, it is possible to construct an
infinite set of othogonal polynomials for the measure%
\begin{equation}
d\mu \left( x;\xi _{i}\right) =\dprod\limits_{i=1}^{L}\left( x-\xi
_{i}\right) \ d\mu \left( x\right)  \tag{10}
\end{equation}%
The integrals\ $I_{N}\ $corresponding to this measure have been calculated
by Brezin and Hikami $\left[ 5\right] .$

Then, Uvarov $\left[ 6\right] $ in 69 and recently Fyodorov and Strahov $%
\left[ 7-8\right] \ $constructed an infinite set of orthogonal polynomials
for the measure%
\begin{equation}
d\mu \left( x;\xi _{i},y_{i}\right) =\frac{\dprod\limits_{i=1}^{L}\left(
x-\xi _{i}\right) }{\dprod\limits_{i=1}^{M}\left( x-y_{i}\right) }\ d\mu
\left( x\right)  \tag{11}
\end{equation}%
if the variables \ $y_{i}$ are chosen away from the real axis.\ In (2003),
Akemann and Vernizzi $\left[ 9\right] $ (see also ref.$\left[ 10\right] $)\
constructed an infinite set of (bi)orthogonal polynomials corresponding to
the measure%
\begin{equation}
d\mu \left( z,z^{\ast };\xi _{i};\eta _{i}^{\ast }\right)
=\dprod\limits_{i=1}^{L_{2}}\left( z^{\ast }-\eta _{i}^{\ast }\right) \ \
\dprod\limits_{i=1}^{L_{1}}\left( z-\xi _{i}\right) \ \ \ d\mu \left(
z,z^{\ast }\right)  \tag{12}
\end{equation}%
where $d\mu \left( z,z^{\ast }\right) $ is a positive Borel measure over the
complex plane.

In the following sections we construct the biorthogonal polynomials\
corresponding to the measure (8).\ We now explain how to calculate the
integrals $I_{N}$\ and $J_{N}$\ from the biorthogonal polynomials. As usual,
we write the expression 
\begin{equation}
\Delta \left( z\right) =\dprod\limits_{i<j}\left( z_{i}-z_{j}\right)
=\left\vert 
\begin{array}{ccc}
\pi _{N-1}\left( z_{1}\right) & ... & \pi _{N-1}\left( z_{N}\right) \\ 
... & ... & ... \\ 
\pi _{0}\left( z_{1}\right) & ... & \pi _{0}\left( z_{N}\right)%
\end{array}%
\right\vert  \tag{13}
\end{equation}%
where the polynomials $\pi _{n}\left( z\right) $ are any set of monic
polynomials. \ If moreover, the polynomials $\pi _{n}\left( z\right) \ $are
(bi)orthogonal, the variables of integration factorize in the integrals $%
\left( 3-4\right) \ $and the (bi)orthogonality can be used to obtain%
\begin{eqnarray}
Z &=&N!\ \dprod\limits_{i=0}^{N-1}h_{i}  \TCItag{$14$} \\
Z\ I_{N} &=&N!\ \dprod\limits_{i=0}^{N-1}\left\Vert q_{i}\right\Vert ^{2} 
\TCItag{$15$}
\end{eqnarray}%
In the general case, the pseudonorms $\left\Vert q_{i}\right\Vert ^{2}$ are
given in section 3; in the special case where $L_{1}=M_{1}$ and $%
L_{2}=M_{2}, $ we have%
\begin{equation}
\left\Vert q_{i}\right\Vert ^{2}=h_{i}\ \frac{D_{i}}{D_{i-1}}\ \ \ \ \ \ \
i\geq 0  \tag{16}
\end{equation}%
where $D_{i}\ $is defined in (19)\ and $D_{-1}$ is given in (108-109);
consequently%
\begin{equation}
I_{N}=\ \ \frac{D_{N-1}}{D_{-1}}  \tag{17}
\end{equation}%
or from~(108-109) 
\begin{equation}
I_{N}=\left( -\right) ^{\frac{M_{1}\left( M_{1}-1\right) }{2}}\left(
-\right) ^{\frac{M_{2}\left( M_{2}-1\right) }{2}}\ \frac{\dprod%
\limits_{i,j=1}^{M_{2}}\left( x_{i}^{\ast }-\eta _{j}^{\ast }\right) \
\dprod\limits_{i,j=1}^{M_{1}}\left( y_{i}-\xi _{j}\right) }{\Delta \left(
x^{\ast }\right) \ \Delta \left( y\right) \ \Delta \left( \eta ^{\ast
}\right) \ \Delta \left( \xi \right) }\ \ D_{N-1}  \tag{18}
\end{equation}%
where $\Delta \left( y\right) $ is the Vandermonde determinant of the
variables $y_{i}$ and $D_{n}\ $is a determinant expressed in terms of four
kernels $K_{n},\ N_{n},\ A_{n},\ N_{n}^{\ast }$ defined in section 2%
\begin{equation}
D_{n}=\left\vert 
\begin{array}{cccccc}
N_{n}\left( \xi _{1},y_{1}\right) & ... & N_{n}\left( \xi
_{M_{1}},y_{1}\right) & A_{n}\left( x_{1}^{\ast },y_{1}\right) & ... & 
A_{n}\left( x_{M_{2}}^{\ast },y_{1}\right) \\ 
... & ... & ... & ... & ... & ... \\ 
N_{n}\left( \xi _{1},y_{M_{1}}\right) & ... & N_{n}\left( \xi
_{M_{1}},y_{M_{1}}\right) & A_{n}\left( x_{1}^{\ast },y_{M_{1}}\right) & ...
& A_{n}\left( x_{M_{2}}^{\ast },y_{M_{1}}\right) \\ 
K_{n}\left( \xi _{1},\eta _{1}^{\ast }\right) & ... & K_{n}\left( \xi
_{M_{1}},\eta _{1}^{\ast }\right) & N_{n}^{\ast }\left( x_{1},\eta
_{1}\right) & ... & N_{n}^{\ast }\left( x_{M_{2}},\eta _{1}\right) \\ 
... & ... & ... & ... & ... & ... \\ 
K_{n}\left( \xi _{1},\eta _{M_{2}}^{\ast }\right) & ... & K_{n}\left( \xi
_{M_{1}},\eta _{M_{2}}^{\ast }\right) & N_{n}^{\ast }\left( x_{1},\eta
_{M_{2}}\right) & ... & N_{n}^{\ast }\left( x_{M_{2}},\eta _{M_{2}}\right)%
\end{array}%
\right\vert  \tag{19}
\end{equation}

\bigskip

Although $N_{n}\left( \xi ,y\right) $ has a single pole at $\xi =y,$ this
pole is cancelled by the prefactor in $I_{N}.\ $Let us mention that this
type of determinant has also been found in Ref. $\left[ 11\right] \ $for the
correlation functions of several hermitian matrices coupled by a chain of
potentials. Finally, the resolvant $J_{N}$ is calculated in section 4 as the
determinant%
\begin{equation}
J_{N}="D_{N-1}"  \tag{20}
\end{equation}%
where$\ "D_{n}"$is%
\begin{equation}
"\det "\left\vert 
\begin{array}{cccccc}
H_{n}\left( \xi _{1},\xi _{1}\right) & ... & N_{n}\left( \xi _{M_{1}},\xi
_{1}\right) & A_{n}\left( \eta _{1}^{\ast },\xi _{1}\right) & ... & 
A_{n}\left( \eta _{M_{2}}^{\ast },\xi _{1}\right) \\ 
... & ... & ... & ... & ... & ... \\ 
N_{n}\left( \xi _{1},\xi _{M_{1}}\right) & ... & H_{n}\left( \xi
_{M_{1}},\xi _{M_{1}}\right) & A_{n}\left( \eta _{1}^{\ast },\xi
_{M_{1}}\right) & ... & A_{n}\left( \eta _{M_{2}}^{\ast },\xi _{M_{1}}\right)
\\ 
K_{n}\left( \xi _{1},\eta _{1}^{\ast }\right) & ... & K_{n}\left( \xi
_{M_{1}},\eta _{1}^{\ast }\right) & H_{n}^{\ast }\left( \eta _{1},\eta
_{1}\right) & ... & N_{n}^{\ast }\left( \eta _{M_{2}},\eta _{1}\right) \\ 
... & ... & ... & ... & ... & ... \\ 
K_{n}\left( \xi _{1},\eta _{M_{2}}^{\ast }\right) & ... & K_{n}\left( \xi
_{M_{1}},\eta _{M_{2}}^{\ast }\right) & N_{n}^{\ast }\left( \eta _{1},\eta
_{M_{2}}\right) & ... & H_{n}^{\ast }\left( \eta _{M_{2}},\eta
_{M_{2}}\right)%
\end{array}%
\right\vert  \tag{21}
\end{equation}

\bigskip

The kernel $H_{n}\left( \xi ,y\right) \ $is defined in (26) and concists in
subtracting the pole of $N_{n}\left( \xi ,y\right) \ $at $\xi =y.$\ The
determinant $"D_{n}"$ is obtained from the determinant $D_{n}$ by changing
all $N_{n}$ on the diagonal into $H_{n},$ changing all $y_{k}$ into $\xi
_{k},\ $all $x_{k}$ into $\eta _{k}$ and finally ignoring all double poles
at $\xi _{j}=\xi _{k}$ and at $\eta _{j}^{\ast }=\eta _{k}^{\ast }\ $as we
develop the determinant (which we denote by " "). In fact, there is no
single poles either at $\xi _{j}=\xi _{k}$ or at $\eta _{j}^{\ast }=\eta
_{k}^{\ast }$ since the residues are zero.

\bigskip

We finally mention that the large $\xi _{i},\ \eta _{k}^{\ast \ }$ expansion
of $J_{N}$\ (which will be done somewhere else $\left[ 12\right] $)\ is
related, in the case of the Gaussian potential, to the number of graphs
obtained from (1)\ by Wick's contraction. The large $N$ expansion of this
number classifies the graphs according to their genus (t'Hooft expansion );
another large $N\ $limit which is considered in $\left[ 12\right] \ $is the
so called BMN limit $\left[ 13-14-15\right] $ where, in (1), $\frac{J_{i}}{%
\sqrt{N}}$ and $\frac{K_{i}}{\sqrt{N}}$ are kept constant.

\bigskip

\bigskip

\bigskip

\section{\protect\bigskip The functions and the kernels}

In this section we describe the biorthogonal polynomials for the four
operations described in the introduction; then we give several simple
examples of measures before describing the general case and the proofs in
section 3.

\bigskip

\qquad \qquad \qquad - operation1:%
\begin{eqnarray}
p_{n}\left( z\right) &\rightarrow &q_{n}\left( z;\xi ,\Phi ;\Phi ,\Phi
\right) =\frac{1}{z-\xi }\ \left\vert 
\begin{array}{cc}
p_{n+1}\left( z\right) & p_{n+1}\left( \xi \right) \\ 
p_{n}\left( z\right) & p_{n}\left( \xi \right)%
\end{array}%
\right\vert \ \ \frac{1}{p_{n}\left( \xi \right) }  \nonumber \\
p_{n}^{\ast }\left( z\right) &\rightarrow &q_{n}^{\ast }\left( z;\Phi ,\xi
^{\ast };\Phi ,\Phi \right) =\ K_{n}^{\ast }\left( z;\xi ^{\ast }\right) \ \ 
\frac{h_{n}}{p_{n}\left( \xi \right) }  \nonumber \\
h_{n} &\rightarrow &\left\Vert q_{n}\right\Vert ^{2}=-h_{n}\ \frac{%
p_{n+1}\left( \xi \right) }{p_{n}\left( \xi \right) }  \TCItag{22}
\end{eqnarray}%
In (22), the symbol $\Phi $ means an empty set of variables and the kernel $%
K_{n}\left( z;\xi ^{\ast }\right) $ is%
\begin{equation}
K_{n}\left( z,\xi ^{\ast }\right) =\sum_{i=0}^{n}\frac{p_{i}\left( z\right)
\ p_{i}^{\ast }\left( \xi \right) }{h_{i}}  \tag{23}
\end{equation}%
The iteration of operation1 to several sources $\xi _{i}$ is the
generalization to potentials of two variables of the Christoffel's
construction with potential of one variable (in the case of one real
variable $x,$ $q_{n}\left( x;\xi \right) $ and $q_{n}^{\ast }\left( x;\xi
\right) $ have similar forms as in (22) and are equal by Christoffel-Darboux
relation; such a relation does not exist with two variables $z$ and $z^{\ast
}$).

\bigskip

\bigskip

\qquad \qquad \qquad - operation2:%
\begin{eqnarray}
p_{n}\left( z\right) &\rightarrow &q_{n}\left( z;\Phi ,\eta ^{\ast };\Phi
,\Phi \right) =\ K_{n}\left( z;\eta ^{\ast }\right) \ \frac{h_{n}}{%
p_{n}^{\ast }\left( \eta \right) }  \nonumber \\
p_{n}^{\ast }\left( z\right) &\rightarrow &q_{n}^{\ast }\left( z;\eta ,\Phi
;\Phi ,\Phi \right) =\frac{1}{z^{\ast }-\eta ^{\ast }}\ \left\vert 
\begin{array}{cc}
p_{n+1}^{\ast }\left( z\right) & p_{n+1}^{\ast }\left( \eta \right) \\ 
p_{n}^{\ast }\left( z\right) & p_{n}^{\ast }\left( \eta \right)%
\end{array}%
\right\vert \ \ \frac{1}{p_{n}^{\ast }\left( \eta \right) }  \nonumber \\
h_{n} &\rightarrow &\left\Vert q_{n}\right\Vert ^{2}=-h_{n}\ \frac{%
p_{n+1}^{\ast }\left( \eta \right) }{p_{n}^{\ast }\left( \eta \right) } 
\TCItag{24}
\end{eqnarray}

\newpage

\qquad \qquad \qquad - operation3:

\bigskip

We introduce the function%
\begin{equation}
t_{n}\left( y\right) =\int d\mu \left( z,z^{\ast }\right) \ p_{n}^{\ast
}\left( z\right) \ \frac{1}{z-y}  \tag{25}
\end{equation}%
which, in the case of potentials of one variable is called the
Cauchy-Hilbert transform of the polynomial $p_{n}\left( x\right) .$ We also
introduce the kernels%
\begin{eqnarray}
H_{n}\left( z,y\right) &=&\sum_{i=0}^{n}\frac{p_{i}\left( z\right) \
t_{i}\left( y\right) }{h_{i}}  \TCItag{26} \\
N_{n}\left( z,y\right) &=&\frac{1}{y-z}+H_{n}\left( z,y\right)  \TCItag{27}
\end{eqnarray}%
Then,%
\begin{eqnarray}
p_{n}\left( z\right) &\rightarrow &q_{n}\left( z;\Phi ,\Phi ;y,\Phi \right)
=\left( z-y\right) \ \ N_{n-1}\left( z,y\right) \ \ \frac{h_{n-1}}{%
t_{n-1}\left( y\right) }\ \ \ \ n>0  \nonumber \\
\ p_{n}^{\ast }\left( z\right) &\rightarrow &q_{n}^{\ast }\left( z;\Phi
,\Phi ;\Phi ,y^{\ast }\right) =\left\vert 
\begin{array}{cc}
\ p_{n}^{\ast }\left( z\right) & t_{n}\left( y\right) \\ 
\ p_{n-1}^{\ast }\left( z\right) & t_{n-1}\left( y\right)%
\end{array}%
\right\vert \ \frac{1}{t_{n-1}\left( y\right) }\ \ \ \ n>0  \nonumber \\
h_{n} &\rightarrow &\left\Vert q_{n}\right\Vert ^{2}=-h_{n-1}\ \frac{%
t_{n}\left( y\right) }{t_{n-1}\left( y\right) }\ \ \ n>0,\ \ \ \ \
\left\Vert q_{0}\right\Vert ^{2}=t_{0}\left( y\right) \   \TCItag{28}
\end{eqnarray}

\bigskip

\qquad \qquad \qquad - operation4:%
\begin{eqnarray}
p_{n}\left( z\right) &\rightarrow &q_{n}\left( z;\Phi ,\Phi ;\Phi ,x^{\ast
}\right) =\left\vert 
\begin{array}{cc}
\ p_{n}\left( z\right) & t_{n}^{\ast }\left( x\right) \\ 
\ p_{n-1}\left( z\right) & t_{n-1}^{\ast }\left( x\right)%
\end{array}%
\right\vert \ \frac{1}{t_{n-1}^{\ast }\left( x\right) }\ \ \ \ n>0  \nonumber
\\
p_{n}^{\ast }\left( z\right) &\rightarrow &q_{n}^{\ast }\left( z;\Phi ,\Phi
;x,\Phi \right) =\left( z^{\ast }-x^{\ast }\right) \ \ N_{n-1}^{\ast }\left(
z,x\right) \ \ \frac{h_{n-1}}{t_{n-1}^{\ast }\left( x\right) }\ \ \ n>0 
\nonumber \\
h_{n} &\rightarrow &\left\Vert q_{n}\right\Vert ^{2}=-h_{n-1}\ \frac{%
t_{n}^{\ast }\left( x\right) }{t_{n-1}^{\ast }\left( x\right) }\ \ \ n>0,\ \
\ \ \ \left\Vert q_{0}\right\Vert ^{2}=t_{0}^{\ast }\left( x\right) 
\TCItag{29}
\end{eqnarray}

\bigskip

We see from these four operations that the biorthogonal polynomials take
different forms according to the measure.\ Before proceeding to the general
case and to the general proofs, let us give some simple examples which make
the general result easier to understand. We introduce the function%
\begin{equation}
Q\left( x^{\ast },y\right) =\int d\mu \left( z,z^{\ast }\right) \ \ \frac{1}{%
\left( z^{\ast }-x^{\ast }\right) \left( z-y\right) }  \tag{30}
\end{equation}%
and the kernel%
\begin{equation}
A_{n}\left( x^{\ast },y\right) =\sum_{i=0}^{n}\frac{t_{i}^{\ast }\left(
x\right) \ t_{i}\left( y\right) }{h_{i}}-Q\left( x^{\ast },y\right)  \tag{31}
\end{equation}

\bigskip

We consider the measure%
\begin{equation}
d\mu \left( z,z^{\ast };\xi ,\eta ^{\ast };y,x^{\ast }\right) =\frac{\left(
z^{\ast }-\eta ^{\ast }\right) }{\left( z^{\ast }-x^{\ast }\right) }\ \frac{%
\left( z-\xi \right) }{\left( z-y\right) }\ \ d\mu \left( z,z^{\ast }\right)
\tag{32}
\end{equation}%
Then, it is easy to find that%
\begin{eqnarray}
p_{n}\left( z\right) &\rightarrow &q_{n}\left( z;\xi ,\eta ^{\ast
};y,x^{\ast }\right)  \nonumber \\
&=&\frac{z-y}{z-\xi }\ \left\vert 
\begin{array}{ccc}
p_{n}\left( z\right) & p_{n}\left( \xi \right) & t_{n}^{\ast }\left( x\right)
\\ 
N_{n}\left( z,y\right) & N_{n}\left( \xi ,y\right) & A_{n}\left( x^{\ast
},y\right) \\ 
K_{n}(z,\eta ^{\ast }) & K_{n}(\xi ,\eta ^{\ast }) & N_{n}^{\ast }\left(
\eta ,x\right)%
\end{array}%
\right\vert \ \ \frac{1}{D_{n-1}}\ \ \ \ \ n>0  \nonumber \\
D_{n} &=&\left\vert 
\begin{array}{cc}
N_{n}\left( \xi ,y\right) & A_{n}\left( x^{\ast },y\right) \\ 
K_{n}(\xi ,\eta ^{\ast }) & N_{n}^{\ast }\left( \eta ,x\right)%
\end{array}%
\right\vert  \nonumber \\
\left\Vert q_{n}\right\Vert ^{2} &=&h_{n}\ \frac{D_{n}}{D_{n-1}}\ \ \ \ \
n>0,\ \ \ \ \ \ \left\Vert q_{0}\right\Vert ^{2}=h_{0}\ \left( x^{\ast
}-\eta ^{\ast }\right) \left( y-\xi \right) \ \ D_{0}  \TCItag{33}
\end{eqnarray}%
We let the reader calculate $q_{n}^{\ast }\left( z;\eta ,\xi ^{\ast
};x,y^{\ast }\right) $ from $q_{n}\left( z;\xi ,\eta ^{\ast };y,x^{\ast
}\right) $ by complex conjugation and by the exchange $\xi \Leftrightarrow
\eta ,\ x\Leftrightarrow y.$

\bigskip

Now, if we let the variable $\eta ^{\ast }\rightarrow \infty \ $in (32-33)$,$
we obtain the biorthogonal polynomials for the measure%
\begin{equation}
d\mu \left( z,z^{\ast };\xi ,\Phi ;y,x^{\ast }\right) =\frac{1}{\left(
z^{\ast }-x^{\ast }\right) }\ \frac{\left( z-\xi \right) }{\left( z-y\right) 
}\ \ d\mu \left( z,z^{\ast }\right)  \tag{34}
\end{equation}%
Performing the corresponding expansion in the kernels, we obtain%
\begin{eqnarray}
p_{n}\left( z\right) &\rightarrow &q_{n}\left( z;\xi ,\Phi ;y,x^{\ast
}\right)  \nonumber \\
&=&\frac{z-y}{z-\xi }\ \left\vert 
\begin{array}{ccc}
p_{n}\left( z\right) & p_{n}\left( \xi \right) & t_{n}^{\ast }\left( x\right)
\\ 
p_{n-1}\left( z\right) & p_{n-1}\left( \xi \right) & t_{n-1}^{\ast }\left(
x\right) \\ 
N_{n}\left( z,y\right) & N_{n}\left( \xi ,y\right) & A_{n}\left( x^{\ast
},y\right)%
\end{array}%
\right\vert \ \ \frac{1}{D_{n-1}}\ \ \ \ \ n>0  \nonumber \\
D_{n} &=&\left\vert 
\begin{array}{cc}
p_{n}\left( \xi \right) & t_{n}^{\ast }\left( x\right) \\ 
N_{n}\left( \xi ,y\right) & A_{n}\left( x^{\ast },y\right)%
\end{array}%
\right\vert \ \ \ \ \ \   \nonumber \\
\left\Vert q_{n}\right\Vert ^{2} &=&-h_{n-1}\ \frac{D_{n}}{D_{n-1}}\ \ \ \ \
n>0,\ \ \ \ \ \ \left\Vert q_{0}\right\Vert ^{2}=-\left( y-\xi \right) \ \
D_{0}  \TCItag{35}
\end{eqnarray}

\bigskip

\bigskip If we let the variable $\xi \rightarrow \infty \ $in (32-33)$,$ we
obtain the biorthogonal polynomials for the measure%
\begin{equation}
d\mu \left( z,z^{\ast };\xi ,\Phi ;y,x^{\ast }\right) =\frac{\left( z^{\ast
}-\eta ^{\ast }\right) }{\left( z^{\ast }-x^{\ast }\right) }\ \frac{1}{%
\left( z-y\right) }\ \ d\mu \left( z,z^{\ast }\right)  \tag{36}
\end{equation}%
\bigskip Performing the corresponding expansion in the kernels, we obtain%
\begin{eqnarray}
p_{n}\left( z\right) &\rightarrow &q_{n}\left( z;\xi ,\Phi ;y,x^{\ast
}\right)  \nonumber \\
&=&\left( z-y\right) \ \left\vert 
\begin{array}{cc}
N_{n-1}\left( z,y\right) & A_{n-1}\left( x^{\ast },y\right) \\ 
K_{n-1}\left( z,\eta ^{\ast }\right) & N_{n-1}^{\ast }\left( \eta ,x\right)%
\end{array}%
\right\vert \ \ \frac{h_{n-1}}{D_{n-1}}\ \ \ \ \ n>0  \nonumber \\
D_{n} &=&\left\vert 
\begin{array}{cc}
t_{n}\left( y\right) & A_{n}\left( x^{\ast },y\right) \\ 
p_{n}^{\ast }\left( \eta \right) & N_{n}^{\ast }\left( \eta ,x\right)%
\end{array}%
\right\vert \ \ \ \ \ \   \nonumber \\
\left\Vert q_{n}\right\Vert ^{2} &=&-h_{n-1}\ \frac{D_{n}}{D_{n-1}}\ \ \ \ \
n>0,\ \ \ \ \ \ \left\Vert q_{0}\right\Vert ^{2}=\left( x^{\ast }-\eta
^{\ast }\right) \ \ D_{0}  \TCItag{37}
\end{eqnarray}

\bigskip

Also, if we let in (32-33) the variable $x^{\ast }\rightarrow \infty $ and
if we use the following result proved in Appendix A%
\begin{eqnarray}
t_{n}\left( y\right) &\sim &-\frac{h_{n}}{y^{n+1}}\ \ \ \ \text{as }%
y\rightarrow \infty  \TCItag{38} \\
N_{n}\left( \xi ,y\right) &\sim &\frac{p_{n+1}\left( \xi \right) }{y^{n+2}}\ 
\text{as }y\rightarrow \infty  \TCItag{39} \\
A_{n}\left( x^{\ast },y\right) &\sim &\frac{t_{n+1}^{\ast }\left( x\right) }{%
y^{n+2}}\ \text{as }y\rightarrow \infty  \TCItag{40}
\end{eqnarray}%
we obtain the biorthogonal polynomials for the measure

\begin{equation}
d\mu \left( z,z^{\ast };\xi ,\eta ^{\ast };y,\Phi \right) =\left( z^{\ast
}-\eta ^{\ast }\right) \ \frac{\left( z-\xi \right) }{\left( z-y\right) }\ \
d\mu \left( z,z^{\ast }\right)  \tag{41}
\end{equation}%
as%
\begin{eqnarray}
p_{n}\left( z\right) &\rightarrow &q_{n}\left( z;\xi ,\eta ^{\ast };y,\Phi
\right)  \nonumber \\
&=&\frac{z-y}{z-\xi }\ \left\vert 
\begin{array}{cc}
N_{n}\left( z,y\right) & N_{n}\left( \xi ,y\right) \\ 
K_{n}\left( z,\eta ^{\ast }\right) & K_{n}\left( \xi ,\eta ^{\ast }\right)%
\end{array}%
\right\vert \ \ \frac{h_{n}}{D_{n-1}}\ \ \ \ \ n>0  \nonumber \\
D_{n} &=&\left\vert 
\begin{array}{cc}
t_{n+1}\left( y\right) & N_{n}\left( \xi ,y\right) \\ 
p_{n+1}^{\ast }\left( \eta \right) & K_{n}\left( \xi ,\eta ^{\ast }\right)%
\end{array}%
\right\vert  \nonumber \\
\left\Vert q_{n}\right\Vert ^{2} &=&-h_{n}\ \frac{D_{n}}{D_{n-1}}\ \ \ \
n>0,\ \ \ \ \ \ \ \ \ \left\Vert q_{0}\right\Vert ^{2}=-h_{0}\ \left( y-\xi
\right) \ D_{0}  \TCItag{42}
\end{eqnarray}%
Finally, if we let in (32-33) the variable $y\rightarrow \infty $ \ we
obtain the biorthogonal polynomials for the measure

\begin{equation}
d\mu \left( z,z^{\ast };\xi ,\eta ^{\ast };y,\Phi \right) =\frac{\left(
z^{\ast }-\eta ^{\ast }\right) }{\left( z^{\ast }-x^{\ast }\right) }\left(
z-\xi \right) \ \ d\mu \left( z,z^{\ast }\right)  \tag{43}
\end{equation}%
as%
\begin{eqnarray}
p_{n}\left( z\right) &\rightarrow &q_{n}\left( z;\xi ,\eta ^{\ast };\Phi
,x^{\ast }\right)  \nonumber \\
&=&\frac{1}{z-\xi }\ \left\vert 
\begin{array}{ccc}
p_{n+1}\left( z\right) & p_{n+1}\left( \xi \right) & t_{n+1}^{\ast }\left(
x\right) \\ 
p_{n}\left( z\right) & p_{n}\left( \xi \right) & t_{n}^{\ast }\left( x\right)
\\ 
K_{n}\left( z,\eta ^{\ast }\right) & K_{n}\left( \xi ,\eta ^{\ast }\right) & 
N_{n}^{\ast }\left( \eta ,x\right)%
\end{array}%
\right\vert \ \ \frac{1}{D_{n-1}}\ \ \ \ \ n>0  \nonumber \\
D_{n} &=&\left\vert 
\begin{array}{cc}
p_{n+1}\left( \xi \right) & t_{n+1}^{\ast }\left( x\right) \\ 
K_{n}\left( \xi ,\eta ^{\ast }\right) & N_{n}^{\ast }\left( \eta ,x\right)%
\end{array}%
\right\vert \ \ \ \ \ n\geq 0  \nonumber \\
\left\Vert q_{n}\right\Vert ^{2} &=&h_{n}\ \frac{D_{n}}{D_{n-1}}\ \ \ \
n>0,\ \ \ \ \ \ \ \ \ \left\Vert q_{0}\right\Vert ^{2}=-h_{0}\ \left(
x^{\ast }-\eta ^{\ast }\right) \ D_{0}  \TCItag{44}
\end{eqnarray}

\bigskip

The rest of this section is devoted to six other simple examples:

1$%
{{}^\circ}%
)$%
\begin{equation}
d\mu \left( z,z^{\ast };\xi ,\Phi ;y,\Phi \right) =\ \frac{\left( z-\xi
\right) }{\left( z-y\right) }\ \ d\mu \left( z,z^{\ast }\right)  \tag{45}
\end{equation}%
which gives%
\begin{eqnarray}
p_{n}\left( z\right) &\rightarrow &q_{n}\left( z;\xi ,\Phi ;y,\Phi \right) 
\nonumber \\
&=&\frac{z-y}{z-\xi }\ \left\vert 
\begin{array}{cc}
p_{n}\left( z\right) & p_{n}\left( \xi \right) \\ 
N_{n}\left( z,y\right) & N_{n}\left( \xi ,y\right)%
\end{array}%
\right\vert \ \ \frac{1}{N_{n-1}\left( \xi ,y\right) }\ \ \ \ \ n>0 
\nonumber \\
\left\Vert q_{n}\right\Vert ^{2} &=&h_{n}\ \frac{N_{n}\left( \xi ,y\right) }{%
N_{n-1}\left( \xi ,y\right) }\ \ \ \ \ n>0,\ \ \ \left\Vert q_{0}\right\Vert
^{2}=h_{0}\ \left( y-\xi \right) \ N_{0}\left( \xi ,y\right)  \TCItag{46}
\end{eqnarray}

2$%
{{}^\circ}%
)$%
\begin{equation}
d\mu \left( z,z^{\ast };\Phi ,\eta ^{\ast };\Phi ,x^{\ast }\right) =\frac{%
\left( z^{\ast }-\eta ^{\ast }\right) }{\left( z^{\ast }-x^{\ast }\right) }\
\ \ d\mu \left( z,z^{\ast }\right)  \tag{47}
\end{equation}

which gives%
\begin{eqnarray}
p_{n}\left( z\right) &\rightarrow &q_{n}\left( z;\xi ,\Phi ;y,\Phi \right) 
\nonumber \\
&=&\frac{z-y}{z-\xi }\ \left\vert 
\begin{array}{cc}
p_{n}\left( z\right) & t_{n}^{\ast }\left( x\right) \\ 
K_{n}\left( z,\eta ^{\ast }\right) & N_{n}^{\ast }\left( \eta ,x\right)%
\end{array}%
\right\vert \ \ \frac{1}{N_{n-1}^{\ast }\left( \eta ,x\right) }\ \ \ \ \ n>0
\nonumber \\
\left\Vert q_{n}\right\Vert ^{2} &=&h_{n}\ \frac{N_{n}^{\ast }\left( \eta
,x\right) }{N_{n-1}^{\ast }\left( \eta ,x\right) }\ \ \ n>0,\ \left\Vert
q_{0}\right\Vert ^{2}=h_{0}\ \left( x^{\ast }-\eta ^{\ast }\right) \
N_{0}^{\ast }\left( \eta ,x\right)  \TCItag{48}
\end{eqnarray}

3$%
{{}^\circ}%
)$\qquad \qquad \qquad 
\begin{equation}
d\mu \left( z,z^{\ast };\xi ,\eta ^{\ast };\Phi ,\Phi \right) =\left(
z^{\ast }-\eta ^{\ast }\right) \left( z-\xi \right) \ \ \ d\mu \left(
z,z^{\ast }\right)  \tag{49}
\end{equation}%
which gives%
\begin{eqnarray}
p_{n}\left( z\right) &\rightarrow &q_{n}\left( z;\xi ,\eta ^{\ast };\Phi
,\Phi \right)  \nonumber \\
&=&\frac{1}{z-\xi }\ \left\vert 
\begin{array}{cc}
p_{n+1}\left( z\right) & p_{n+1}\left( \xi \right) \\ 
K_{n}\left( z,\eta ^{\ast }\right) & K_{n}\left( \xi ,\eta ^{\ast }\right)%
\end{array}%
\right\vert \ \ \frac{1}{K_{n}\left( \xi ,\eta ^{\ast }\right) }\ \ \ n\geq 0
\nonumber \\
\left\Vert q_{n}\right\Vert ^{2} &=&h_{n+1}\ \frac{K_{n+1}\left( \xi ,\eta
^{\ast }\right) }{K_{n}\left( \xi ,\eta ^{\ast }\right) }\ \ \ \ \ n\geq 0 
\TCItag{50}
\end{eqnarray}

4$%
{{}^\circ}%
)$%
\begin{equation}
d\mu \left( z,z^{\ast };\Phi ,\Phi ;y,x^{\ast }\right) =\frac{1}{\left(
z^{\ast }-x^{\ast }\right) \left( z-y\right) }\ \ \ d\mu \left( z,z^{\ast
}\right)  \tag{51}
\end{equation}%
which gives%
\begin{eqnarray}
p_{n}\left( z\right) &\rightarrow &q_{n}\left( z;\Phi ,\Phi ;y,x^{\ast
}\right)  \nonumber \\
&=&\left( z-y\right) \ \left\vert 
\begin{array}{cc}
p_{n-1}\left( z\right) & t_{n-1}^{\ast }\left( x\right) \\ 
N_{n-1}\left( z,y\right) & A_{n-1}\left( x^{\ast },y\right)%
\end{array}%
\right\vert \ \frac{1}{D_{n-1}}\ \ \ \ \ \ \ n>0  \nonumber \\
D_{n} &=&A_{n-1}\left( x^{\ast },y\right) \ \ \ n>0,\ \ \ \ \ \
D_{0}=-Q\left( x^{\ast },y\right)  \nonumber \\
\left\Vert q_{n}\right\Vert ^{2} &=&h_{n-1}\ \frac{D_{n}}{D_{n-1}}\ \ \ \
n>0,\ \ \ \ \ \ \ \ \left\Vert q_{0}\right\Vert ^{2}=-D_{0}  \TCItag{52}
\end{eqnarray}

5$%
{{}^\circ}%
)$%
\begin{equation}
d\mu \left( z,z^{\ast };\xi ,\Phi ;\Phi ,x^{\ast }\right) =\frac{\left(
z-\xi \right) }{\left( z^{\ast }-x^{\ast }\right) }\ \ \ d\mu \left(
z,z^{\ast }\right)  \tag{53}
\end{equation}%
which gives%
\begin{eqnarray}
p_{n}\left( z\right) &\rightarrow &q_{n}\left( z;\xi ,\Phi ;\Phi ,x^{\ast
}\right)  \nonumber \\
&=&\frac{1}{z-\xi }\left\vert 
\begin{array}{ccc}
p_{n+1}\left( z\right) & p_{n+1}\left( \xi \right) & t_{n+1}^{\ast }\left(
x\right) \\ 
p_{n}\left( z\right) & p_{n}\left( \xi \right) & t_{n}^{\ast }\left( x\right)
\\ 
p_{n-1}\left( z\right) & p_{n-1}\left( \xi \right) & t_{n-1}^{\ast }\left(
x\right)%
\end{array}%
\right\vert \ \ \frac{1}{D_{n-1}}\ \ \ \ \ \ n>0  \nonumber \\
D_{n} &=&\left\vert 
\begin{array}{cc}
p_{n+1}\left( \xi \right) & t_{n+1}^{\ast }\left( x\right) \\ 
p_{n}\left( \xi \right) & t_{n}^{\ast }\left( x\right)%
\end{array}%
\right\vert  \nonumber \\
\left\Vert q_{n}\right\Vert ^{2} &=&h_{n-1}\ \frac{D_{n}}{D_{n-1}}\ \ \ \
n>0,\ \ \ \ \ \ \ \left\Vert q_{0}\right\Vert ^{2}=-D_{0}  \TCItag{54}
\end{eqnarray}

\bigskip

6$%
{{}^\circ}%
)$%
\begin{equation}
d\mu \left( z,z^{\ast };\Phi ,\eta ^{\ast };y,\Phi \right) =\frac{\left(
z^{\ast }-\eta ^{\ast }\right) }{\left( z-y\right) }\ \ \ d\mu \left(
z,z^{\ast }\right)  \tag{55}
\end{equation}%
which gives%
\begin{eqnarray}
p_{n}\left( z\right) &\rightarrow &q_{n}\left( z;\xi ,\Phi ;\Phi ,x^{\ast
}\right)  \nonumber \\
&=&\left( z-y\right) \left\vert 
\begin{array}{cc}
N_{n-1}\left( z,y\right) & t_{n}\left( y\right) \\ 
K_{n-1}\left( z,\eta ^{\ast }\right) & p_{n}^{\ast }\left( \eta \right)%
\end{array}%
\right\vert \ \ \frac{h_{n-1}}{D_{n-1}}\ \ \ \ \ \ n>0  \nonumber \\
D_{n} &=&\left\vert 
\begin{array}{cc}
t_{n}\left( y\right) & t_{n+1}\left( y\right) \\ 
p_{n}^{\ast }\left( \eta \right) & p_{n+1}^{\ast }\left( \eta \right)%
\end{array}%
\right\vert \ \ n\geq 0  \nonumber \\
\left\Vert q_{n}\right\Vert ^{2} &=&-h_{n-1}\ \frac{D_{n}}{D_{n-1}}\ \ \ \
n>0,\ \ \ \ \ \ \ \left\Vert q_{0}\right\Vert ^{2}=-D_{0}  \TCItag{56}
\end{eqnarray}

\bigskip

\bigskip

\bigskip

\section{\protect\bigskip The biorthogonal polynomials}

\bigskip

We now consider the measure $\left( 8\right) $. We learned from the previous
section how to construct the biorthogonal polynomials; clearly, it is of the
form%
\begin{equation}
q_{n}\left( z;\xi _{i},\eta _{i}^{\ast };y_{i},x_{i}^{\ast }\right) =\frac{%
\dprod\limits_{i=1}^{M_{1}}\left( z-y_{i}\right) }{\dprod%
\limits_{i=1}^{L_{1}}\left( z-\xi _{i}\right) }\ \ \left\vert 
\begin{array}{ccc}
. & . & . \\ 
. & . & . \\ 
. & . & .%
\end{array}%
\right\vert _{d,\gamma }\ \ \ \frac{a_{d,\gamma }}{D_{d,\gamma }}  \tag{57}
\end{equation}%
where $\frac{a_{d,\gamma }}{D_{d,\gamma }}$ is a factor which makes the
polynomial $q_{n}\ $monic. The form of the determinant depends of two
parameters%
\begin{eqnarray}
d &=&n+L_{1}-M_{1}  \TCItag{58} \\
\gamma &=&n+L_{2}-M_{2}  \TCItag{59}
\end{eqnarray}%
and its construction is always related to the linear properties of its
columns.

Clearly,\ the expression$\ \dprod\limits_{i=1}^{M_{1}}\left( z-y_{i}\right)
\ast \left\vert _{.}\right\vert _{d,\gamma }\ $must be a polynomial in $z$
of degree $\left( n+L_{1}\right) \ $which vanishes when $z=\xi _{i}\ $for
any $i.$ In the determinant, the first column is the only $z$ dependant
column; then, some of the $L_{1}\ $columns at the right of the first one are
chosen to be identical to the first one with $z$ successively replaced by $%
\xi _{1},...,\xi _{L_{1}}.$ Consequently, the polynomial $%
\dprod\limits_{i=1}^{M_{1}}\left( z-y_{i}\right) \ast \ \left\vert
.\right\vert _{d,\gamma }$ vanishes for $z=\xi _{i}$ for any $i,$ and can be
divided by $\dprod\limits_{i=1}^{L_{1}}\left( z-\xi _{i}\right) $ to give a
polynomial of degree $n$.\ 

\ If $d\geq 0\ $\ we may introduce in the first column the polynomial $%
p_{d}\left( z\right) ,\ $the kernels $K_{d}\left( z,\eta ^{\ast }\right) ,\ $
$N_{d}\left( z,y\right) .\ $If $\ d<0\ $it is convenient to introduce%
\begin{eqnarray}
N_{k<0}\left( z,y\right) &=&-\frac{1}{z-y}  \TCItag{60} \\
K_{k<0}\left( z,\eta ^{\ast }\right) &=&0  \TCItag{61} \\
A_{k<0}\left( x^{\ast },y\right) &=&-Q\left( x^{\ast },y\right)  \TCItag{62}
\end{eqnarray}%
so that if $d=-1\ $the expression $\dprod\limits_{i=1}^{M_{1}}\left(
z-y_{i}\right) \ N_{-1}\left( z,y_{k}\right) \ $is a polynomial in $z$ of
degree $\left( n+L_{1}\right) .\ $Finally, if $d<-1\ $we introduce in the
determinant a line 
\begin{equation}
\left\vert 
\begin{array}{cccc}
N_{-1}\left( z,y\right) & p_{0}\left( y\right) & ... & p_{-d-2}\left(
y\right)%
\end{array}%
\right\vert \sim z^{d}\ \ \ \ \text{as \ }z\rightarrow \infty  \tag{63}
\end{equation}

\bigskip

\ The remaining columns are constructed in such a way that%
\begin{equation}
\int d\mu \left( z,z^{\ast };\xi _{i},\eta _{i}^{\ast };y_{i},x_{i}^{\ast
}\right) \ p_{m}^{\ast }\left( z\right) \ q_{n}\left( z;\xi _{i},\eta
_{i}^{\ast };y_{i},x_{i}^{\ast }\right) \ =0\ \ \ \ \ \ \text{for }m<n\  
\tag{64}
\end{equation}%
or equivalently%
\begin{equation}
\int d^{2}z\ p_{m}^{\ast }\left( z\right) \ \frac{\dprod%
\limits_{i=1}^{L_{2}}\left( z^{\ast }-\eta _{i}^{\ast }\right) }{%
\dprod\limits_{i=1}^{M_{2}}\left( z^{\ast }-x_{i}^{\ast }\right) }\ \
\left\vert 
\begin{array}{ccc}
. & . & . \\ 
. & . & . \\ 
. & . & .%
\end{array}%
\right\vert _{\ d,\gamma }\ e^{-V\left( z,z^{\ast }\right) }=0\ \ \ \ \ \ \ 
\text{for }m<n  \tag{65}
\end{equation}%
The square pseudonorm is calculated by 
\begin{equation}
\int d^{2}z\ p_{n}^{\ast }\left( z\right) \ \frac{\dprod%
\limits_{i=1}^{L_{2}}\left( z^{\ast }-\eta _{i}^{\ast }\right) }{%
\dprod\limits_{i=1}^{M_{2}}\left( z^{\ast }-x_{i}^{\ast }\right) }\ \
\left\vert 
\begin{array}{ccc}
. & . & . \\ 
. & . & . \\ 
. & . & .%
\end{array}%
\right\vert _{\ d,\gamma }\ e^{-V\left( z,z^{\ast }\right) }=\left\Vert
q_{n}\right\Vert ^{2}  \tag{66}
\end{equation}

\bigskip

\bigskip

We use the property 
\begin{eqnarray}
&&p_{m}^{\ast }\left( z\right) \ \frac{\dprod\limits_{i=1}^{L_{2}}\left(
z^{\ast }-\eta _{i}^{\ast }\right) }{\dprod\limits_{i=1}^{M_{2}}\left(
z^{\ast }-x_{i}^{\ast }\right) }  \nonumber \\
&=&\pi _{m+L_{2}-M_{2}}\left( z^{\ast }\right) +\frac{1}{\Delta \left(
x_{i}^{\ast }\right) }\left\vert 
\begin{array}{ccc}
\frac{p_{m}^{\ast }\left( x_{1}\right) \varphi _{1}\left( \eta _{i}^{\ast
}\right) }{z^{\ast }-x_{1}^{\ast }} & ... & \frac{p_{m}^{\ast }\left(
x_{M_{2}}\right) \varphi _{M_{2}}\left( \eta _{i}^{\ast }\right) }{z^{\ast
}-x_{M_{2}}^{\ast }} \\ 
p_{M_{2}-2}^{\ast }\left( x_{1}\right) & ... & p_{M_{2}-2}^{\ast }\left(
x_{M_{2}}\right) \\ 
... & ... & ... \\ 
p_{0}^{\ast }\left( x_{1}\right) & ... & p_{0}^{\ast }\left( x_{M_{2}}\right)%
\end{array}%
\right\vert  \TCItag{67}
\end{eqnarray}%
where%
\begin{equation}
\varphi _{j}\left( \eta _{i}^{\ast }\right)
=\dprod\limits_{i=1}^{L_{2}}\left( x_{j}^{\ast }-\eta _{i}^{\ast }\right) 
\tag{68}
\end{equation}%
and where the polynomial 
\begin{equation}
\pi _{m+L_{2}-M_{2}}\left( z^{\ast }\right)
=\sum_{i=0}^{m+L_{2}-M_{2}}\alpha _{i}\ p_{i}^{\ast }\left( z\right) \ \ \
,\ \ \ \ \ \ \ \ \ \alpha _{m+L_{2}-M_{2}}=1  \tag{69}
\end{equation}%
is monic of degree $\left( m+L_{2}-M_{2}\right) $ in $z^{\ast }$ and is$\ 0$
if $\left( m+L_{2}-M_{2}\right) <0.$

\bigskip

If the first column of the determinant contains a term $p_{k}\left( z\right)
,\ $the integrals (65-66) generate a term%
\begin{eqnarray}
p_{k}\left( z\right) &\Rightarrow &\left\{ 
\begin{array}{c}
\alpha _{k}\ h_{k} \\ 
0%
\end{array}%
\right\} +\frac{1}{\Delta \left( x_{i}^{\ast }\right) }  \nonumber \\
&&\left\vert 
\begin{array}{ccc}
p_{m}^{\ast }\left( x_{1}\right) \varphi _{1}\left( \eta _{i}^{\ast }\right)
t_{k}^{\ast }\left( x_{1}\right) & ... & p_{m}^{\ast }\left(
x_{M_{2}}\right) \varphi _{M_{2}}\left( \eta _{i}^{\ast }\right) t_{k}^{\ast
}\left( x_{M_{2}}\right) \\ 
p_{M_{2}-2}^{\ast }\left( x_{1}\right) & ... & p_{M_{2}-2}^{\ast }\left(
x_{M_{2}}\right) \\ 
... & ... & ... \\ 
p_{0}^{\ast }\left( x_{1}\right) & ... & p_{0}^{\ast }\left( x_{M_{2}}\right)%
\end{array}%
\right\vert  \nonumber \\
&&  \TCItag{70}
\end{eqnarray}%
where $\ \alpha _{k}\ h_{k}$ requires $0\leq k\leq \left(
m+L_{2}-M_{2}\right) $ and $0$ otherwise.\ Similarly,\ the integrals (65-66)
generate the terms%
\begin{eqnarray}
N_{k}\left( z,y\right) &\Rightarrow &-\sum_{i=Sup\left( 0,k+1\right)
}^{m+L_{2}-M_{2}}\alpha _{i}\ t_{i}\left( y\right) +\frac{1}{\Delta \left(
x_{i}^{\ast }\right) }  \nonumber \\
&&\left\vert 
\begin{array}{ccc}
p_{m}^{\ast }\left( x_{1}\right) \varphi _{1}\left( \eta _{i}^{\ast }\right)
A_{k}\left( x_{1}^{\ast },y\right) & ... & p_{m}^{\ast }\left(
x_{M_{2}}\right) \varphi _{M_{2}}\left( \eta _{i}^{\ast }\right) A_{k}\left(
x_{M_{2}}^{\ast },y\right) \\ 
p_{M_{2}-2}^{\ast }\left( x_{1}\right) & ... & p_{M_{2}-2}^{\ast }\left(
x_{M_{2}}\right) \\ 
... & ... & ... \\ 
p_{0}^{\ast }\left( x_{1}\right) & ... & p_{0}^{\ast }\left( x_{M_{2}}\right)%
\end{array}%
\right\vert \ \ \ \ \ \ \   \TCItag{71} \\
K_{k}\left( z,\eta _{l}^{\ast }\right) &\Rightarrow &-\sum_{i=Sup\left(
0,k+1\right) }^{m+L_{2}-M_{2}}\alpha _{i}\ p_{i}^{\ast }\left( \eta
_{l}\right) +\frac{1}{\Delta \left( x_{i}^{\ast }\right) }  \nonumber \\
&&\left\vert 
\begin{array}{ccc}
p_{m}^{\ast }\left( x_{1}\right) \varphi _{1}\left( \eta _{i}^{\ast }\right)
N_{k}^{\ast }\left( \eta _{l},x_{1},\right) & ... & p_{m}^{\ast }\left(
x_{M_{2}}\right) \varphi _{M_{2}}\left( \eta _{i}^{\ast }\right) N_{k}^{\ast
}\left( \eta _{l},x_{M_{2}}\right) \\ 
p_{M_{2}-2}^{\ast }\left( x_{1}\right) & ... & p_{M_{2}-2}^{\ast }\left(
x_{M_{2}}\right) \\ 
... & ... & ... \\ 
p_{0}^{\ast }\left( x_{1}\right) & ... & p_{0}^{\ast }\left( x_{M_{2}}\right)%
\end{array}%
\right\vert  \nonumber \\
&&  \TCItag{72}
\end{eqnarray}%
where the sum $\sum $ is absent if $k\geq \left( m+L_{2}-M_{2}\right) $ or
if $\left( m+L_{2}-M_{2}\right) <0.\ $The equations (71-72) are still valid
if $k<0.$

\bigskip

\subsection{$\mathbf{\protect\gamma =n+L}_{2}\mathbf{-M}_{2}\mathbf{<0}$}

Since, in that case, there is no polynomial $\pi _{m+L_{2}-M_{2}}\left(
z^{\ast }\right) $ in (67) for $m\leq n,\ $the calculation of (65-66) as
given by (70-71-72), may be written%
\begin{equation}
\left\{ 
\begin{array}{c}
p_{k}\left( z\right) \\ 
N_{k}\left( z,y\right) \\ 
K_{k}\left( z,\eta _{k}^{\ast }\right)%
\end{array}%
\right\} \Rightarrow \sum_{i=1}^{M_{2}}\beta _{i}\ \left\{ 
\begin{array}{c}
t_{k}^{\ast }\left( x_{i}\right) \\ 
A_{k}\left( x_{i}^{\ast },y\right) \\ 
N_{k}^{\ast }\left( \eta _{k},x_{i}\right)%
\end{array}%
\right\}  \tag{73}
\end{equation}%
where the coefficients $\beta _{i}$ are the same in the three cases.
Consequently, a determinant containing%
\begin{equation}
\ \left\vert 
\begin{array}{cc}
p_{k}\left( z\right) & t_{k}^{\ast }\left( x_{i}\right) \\ 
N_{k}\left( z,y\right) & A_{k}\left( x_{i}^{\ast },y\right) \\ 
K_{k}\left( z,\eta _{k}^{\ast }\right) & N_{k}^{\ast }\left( \eta
_{k},x_{i}\right)%
\end{array}%
\right\vert  \tag{74}
\end{equation}%
gives zero in (65) and (66) for $m\leq n.\ $Now, we observe that the
coefficients $\beta _{i}$ are such that%
\begin{equation}
\sum_{i=1}^{M_{2}}\beta _{i}\ \left\{ 
\begin{array}{c}
p_{c}^{\ast }\left( x_{i}\right) \\ 
p_{c}^{\ast }\left( x_{i}\right)%
\end{array}%
\right\} =\left\{ 
\begin{array}{c}
0\ \ \ \ \ \text{if\ }c<-\left( m+L_{2}-M_{2}+1\right) \\ 
1\ \ \ \ \ \ \text{if\ }c=-\left( m+L_{2}-M_{2}+1\right)%
\end{array}%
\right\}  \tag{75}
\end{equation}

\bigskip $\ \ $

\bigskip

\qquad a$%
{{}^\circ}%
)\ \ \ $if $d\geq 0,$ then$\ q_{n}\left( z;\xi _{i},\eta _{i}^{\ast
};y_{i},x_{i}^{\ast }\right) \ $contains a $\ $determinant of size$\ $

$\left( L_{1}+M_{2}+1\right) \times \left( L_{2}+M_{1}+d-\gamma +1\right) $
\ \ 
\begin{equation}
q_{n}\left( z;\xi _{i},\eta _{i}^{\ast };y_{i},x_{i}^{\ast }\right) =\frac{%
\dprod\limits_{i=1}^{M_{1}}\left( z-y_{i}\right) }{\dprod%
\limits_{i=1}^{L_{1}}\left( z-\xi _{i}\right) }\left\vert 
\begin{array}{ccc}
p_{d}\left( z\right) & p_{d}\left( \xi _{j}\right) & t_{d}^{\ast }\left(
x_{l}\right) \\ 
... & ... & ... \\ 
p_{0}\left( z\right) & p_{0}\left( \xi _{j}\right) & t_{0}^{\ast }\left(
x_{l}\right) \\ 
0 & 0 & p_{0}^{\ast }\left( x_{l}\right) \\ 
... & ... & ... \\ 
0 & 0 & p_{-\gamma -1}^{\ast }\left( x_{l}\right) \\ 
N_{d}\left( z,y_{i}\right) & N_{d}\left( \xi _{j},y_{i}\right) & A_{d}\left(
x_{l}^{\ast },y_{i}\right) \\ 
K_{d}\left( z,\eta _{k}^{\ast }\right) & K_{d}\left( \xi _{j},\eta
_{k}^{\ast }\right) & N_{d}^{\ast }\left( \eta _{k},x_{l}\right)%
\end{array}%
\right\vert \ \frac{1}{D_{d,\gamma }}  \tag{76}
\end{equation}%
where\ $D_{d,\gamma }$ is written in Appendix B ((B1)\ for$\ d>0$ and (B2)\
for$\ d=0).$ The pseudonorm is generated in (76) from the terms $p_{-\gamma
-1}^{\ast }\left( x_{l}\right) ;$ from (75) we find%
\begin{equation}
\left\Vert q_{n}\right\Vert ^{2}=\left( -\right) ^{d-\gamma -1}\ \frac{%
D_{d+1,\gamma +1}}{D_{d,\gamma }}  \tag{77}
\end{equation}%
where $D_{d+1,0}\ $is given in (B4).

\bigskip

\qquad b$%
{{}^\circ}%
)\ \ $if $d=-1,\ $then$\ q_{n}\left( z;\xi _{i},\eta _{i}^{\ast
};y_{i},x_{i}^{\ast }\right) \ $contains a $\ $determinant of size

$\left( L_{1}+M_{2}+1\right) \times \left( L_{2}+M_{1}-\gamma \right) $ 
\begin{eqnarray}
q_{n}\left( z;\xi _{i},\eta _{i}^{\ast };y_{i},x_{i}^{\ast }\right) &=&\frac{%
\dprod\limits_{i=1}^{M_{1}}\left( z-y_{i}\right) }{\dprod%
\limits_{i=1}^{L_{1}}\left( z-\xi _{i}\right) }\ \ \frac{-1}{D_{-1,\gamma }}
\nonumber \\
&&\left\vert 
\begin{array}{ccc}
0 & 0 & p_{0}^{\ast }\left( x_{l}\right) \\ 
... & ... & ... \\ 
0 & 0 & p_{-\gamma -1}^{\ast }\left( x_{l}\right) \\ 
N_{-1}\left( z,y_{i}\right) & N_{-1}\left( \xi _{j},y_{i}\right) & 
A_{-1}\left( x_{l}^{\ast },y_{i}\right) \\ 
0 & 0 & N_{-1}^{\ast }\left( \eta _{k},x_{l}\right)%
\end{array}%
\right\vert \   \TCItag{78}
\end{eqnarray}%
where $D_{-1,\gamma }$ is given in (B3). The pseudonorm is\ stll obtained
from the term$\ p_{-\gamma -1}^{\ast }\left( x_{l}\right) $ as%
\begin{equation}
\left\Vert q_{n}\right\Vert ^{2}=\left( -\right) ^{1-\gamma }\ \frac{%
D_{0,\gamma +1}}{D_{-1,\gamma }}  \tag{79}
\end{equation}%
where $D_{0,\gamma +1}\ $is given in (B2)\ if\ $\gamma <-1\ $and in (B5)\ if$%
\ \gamma =-1.$

\bigskip

\qquad c$%
{{}^\circ}%
)\ $if\ $d<-1,\ $then$\ q_{n}\left( z;\xi _{i},\eta _{i}^{\ast
};y_{i},x_{i}^{\ast }\right) \ $contains a\ determinant$\ $of size

$\ \left( L_{1}+M_{2}-d\right) \times \left( L_{2}+M_{1}-\gamma \right) $%
\begin{eqnarray}
&&q_{n}\left( z;\xi _{i},\eta _{i}^{\ast };y_{i},x_{i}^{\ast }\right) 
\nonumber \\
&=&\frac{\dprod\limits_{i=1}^{M_{1}}\left( z-y_{i}\right) }{%
\dprod\limits_{i=1}^{L_{1}}\left( z-\xi _{i}\right) }\ \ \frac{\left(
-\right) ^{-d}}{D_{d,\gamma }}  \nonumber \\
&&\left\vert 
\begin{array}{cccccc}
0 & 0 & ... & 0 & 0 & p_{0}^{\ast }\left( x_{l}\right) \\ 
... & ... & ... & ... & ... & ... \\ 
0 & 0 & ... & 0 & 0 & p_{-\gamma -1}^{\ast }\left( x_{l}\right) \\ 
N_{-1}\left( z,y_{i}\right) & p_{0}\left( y_{i}\right) & ... & 
p_{-d-2}\left( y_{i}\right) & N_{-1}\left( \xi _{j},y_{i}\right) & 
A_{-1}\left( x_{l}^{\ast },y_{i}\right) \\ 
0 & 0 & ... & 0 & 0 & N_{-1}^{\ast }\left( \eta _{k},x_{l}\right)%
\end{array}%
\right\vert  \nonumber \\
&&  \TCItag{80}
\end{eqnarray}%
where$\ D_{d,\gamma }$\ is given in (B3) The pseudo norm is still obtained
from the term

$\ p_{-\gamma -1}^{\ast }\left( x_{l}\right) \ $as 
\begin{equation}
\left\Vert q_{n}\right\Vert ^{2}=\left( -\right) ^{d-\gamma }\ \frac{%
D_{d+1,\gamma +1}}{D_{d,\gamma }}  \tag{81}
\end{equation}%
where $D_{d+1,0}\ $is given in (B8).\ 

\bigskip

\bigskip

\subsection{$\mathbf{\protect\gamma =n+L}_{2}\mathbf{-M}_{2}\mathbf{\geq 0}$}

\bigskip

\qquad a$%
{{}^\circ}%
)\ $we suppose $d\geq \gamma .$\bigskip

In this case, for $m\leq n,$ equations (65-66) give from (70-71-72)\ 
\begin{equation}
\left\{ 
\begin{array}{c}
p_{k}\left( z\right) \\ 
N_{d}\left( z,y\right) \\ 
K_{d}\left( z,\eta _{k}^{\ast }\right)%
\end{array}%
\right\} \Rightarrow \sum_{i=1}^{M_{2}}\beta _{i}\ \left\{ 
\begin{array}{c}
t_{k}^{\ast }\left( x_{i}\right) \\ 
A_{d}\left( x_{i}^{\ast },y\right) \\ 
N_{d}^{\ast }\left( \eta _{k},x_{i}\right)%
\end{array}%
\right\} \ \ \ \ \ \ \ \ \ \ \ k\geq \gamma +1  \tag{82}
\end{equation}%
while these equations give for $k=\gamma $ 
\begin{equation}
p_{\gamma }\left( z\right) \Rightarrow \left\{ 
\begin{array}{c}
\sum_{i=1}^{M_{2}}\beta _{i}\ t_{\gamma }^{\ast }\left( x_{i}\right) \ \ \ \
\ \ \ \ \ \ \ \ \ \ m<n \\ 
h_{\gamma }+\sum_{i=1}^{M_{2}}\beta _{i}\ t_{\gamma }^{\ast }\left(
x_{i}\right) \ \ \ \ \ \ m=n%
\end{array}%
\right\}  \tag{83}
\end{equation}%
Consequently, the polynomial is a determinant\ of size

$\left( L_{1}+M_{2}+1\right) \times \left( L_{2}+M_{1}+d-\gamma +1\right) $%
\begin{equation}
q_{n}\left( z;\xi _{i},\eta _{i}^{\ast };y_{i},x_{i}^{\ast }\right) =\frac{%
\dprod\limits_{i=1}^{M_{1}}\left( z-y_{i}\right) }{\dprod%
\limits_{i=1}^{L_{1}}\left( z-\xi _{i}\right) }\left\vert 
\begin{array}{ccc}
p_{d}\left( z\right) & p_{d}\left( \xi _{j}\right) & t_{d}^{\ast }\left(
x_{l}\right) \\ 
... & ... & ... \\ 
p_{\gamma }\left( z\right) & p_{\gamma }\left( \xi _{j}\right) & t_{\gamma
}^{\ast }\left( x_{l}\right) \\ 
N_{d}\left( z,y_{i}\right) & N_{d}\left( \xi _{j},y_{i}\right) & A_{d}\left(
x_{l}^{\ast },y_{i}\right) \\ 
K_{d}\left( z,\eta _{k}^{\ast }\right) & K_{d}\left( \xi _{j},\eta
_{k}^{\ast }\right) & N_{d}^{\ast }\left( \eta _{k},x_{l}\right)%
\end{array}%
\right\vert \ \frac{1}{D_{d,\gamma }}  \tag{84}
\end{equation}%
where$\ D_{d,\gamma }\ $is given in (B4)\ if $d>\gamma $.and in (B5) if$\
d=\gamma .$ The pseudonorm comes from the term $p_{\gamma }\left( z\right) $
and is found from (83) to be%
\begin{equation}
\left\Vert q_{n}\right\Vert ^{2}=\left( -\right) ^{d-\gamma }\ h_{\gamma }\ 
\frac{D_{d+1,\gamma +1}}{D_{d,\gamma }}  \tag{85}
\end{equation}%
where $D_{d+1,\gamma +1}$ is also given in (B4)\ if $d>\gamma $.and in (B5)
if$\ d=\gamma .$

\bigskip

b$%
{{}^\circ}%
)\ $we suppose $d=\gamma -1$

\bigskip

In this case, for $m<n$,\ we still have in (65) 
\begin{equation}
\left\{ 
\begin{array}{c}
N_{d}\left( z,y\right) \\ 
K_{d}\left( z,\eta _{k}^{\ast }\right)%
\end{array}%
\right\} \Rightarrow \sum_{i=1}^{M_{2}}\beta _{i}\ \left\{ 
\begin{array}{c}
A_{d}\left( x_{i}^{\ast },y\right) \\ 
N_{d}^{\ast }\left( \eta _{k},x_{i}\right)%
\end{array}%
\right\}  \tag{86}
\end{equation}%
however, for $m=n\ $%
\begin{equation}
\left\{ 
\begin{array}{c}
N_{d}\left( z,y\right) \\ 
K_{d}\left( z,\eta _{k}^{\ast }\right)%
\end{array}%
\right\} \Rightarrow -\left\{ 
\begin{array}{c}
t_{\gamma }\left( y\right) \\ 
p_{\gamma }^{\ast }\left( \eta \right)%
\end{array}%
\right\} +\sum_{i=1}^{M_{2}}\beta _{i}\ \left\{ 
\begin{array}{c}
A_{d}\left( x_{i}^{\ast },y\right) \\ 
N_{d}^{\ast }\left( \eta _{k},x_{i}\right)%
\end{array}%
\right\}  \tag{87}
\end{equation}%
Consequently, the polynomials contain a determinant of size

$\left( L_{1}+M_{2}+1\right) \times \left( L_{2}+M_{1}\right) $%
\begin{eqnarray}
&&q_{n}\left( z;\xi _{i},\eta _{i}^{\ast };y_{i},x_{i}^{\ast }\right) 
\nonumber \\
&=&\frac{\dprod\limits_{i=1}^{M_{1}}\left( z-y_{i}\right) }{%
\dprod\limits_{i=1}^{L_{1}}\left( z-\xi _{i}\right) }\left\vert 
\begin{array}{ccc}
N_{d}\left( z,y_{i}\right) & N_{d}\left( \xi _{j},y_{i}\right) & A_{d}\left(
x_{l}^{\ast },y_{i}\right) \\ 
K_{d}\left( z,\eta _{k}^{\ast }\right) & K_{d}\left( \xi _{j},\eta
_{k}^{\ast }\right) & N_{d}^{\ast }\left( \eta _{k},x_{l}\right)%
\end{array}%
\right\vert \frac{h_{d}}{D_{d,\gamma }}\   \nonumber \\
\text{for}\ \ \ d &>&-1  \TCItag{88} \\
&&q_{n}\left( z;\xi _{i},\eta _{i}^{\ast };y_{i},x_{i}^{\ast }\right) 
\nonumber \\
&=&\frac{\dprod\limits_{i=1}^{M_{1}}\left( z-y_{i}\right) }{%
\dprod\limits_{i=1}^{L_{1}}\left( z-\xi _{i}\right) }\left\vert 
\begin{array}{ccc}
N_{-1}\left( z,y_{i}\right) & N_{-1}\left( \xi _{j},y_{i}\right) & 
A_{-1}\left( x_{l}^{\ast },y_{i}\right) \\ 
0 & 0 & N_{-1}^{\ast }\left( \eta _{k},x_{l}\right)%
\end{array}%
\right\vert \ \frac{\left( -1\right) }{D_{-1,0}}\ \   \nonumber \\
\ \ \ \text{for\ \ \ \ }d &=&-1  \TCItag{89}
\end{eqnarray}%
where$\ D_{d,\gamma }\ $is given in (B6)\ if$\ d>-1$ and $D_{-1,0}$ is given
in (B8). The pseudonorm is obtained from (87) as%
\begin{eqnarray}
\left\Vert q_{n}\right\Vert ^{2} &=&-\ h_{d}\ \frac{D_{d+1,\gamma +1}}{%
D_{d,\gamma }}\ \ \ \ \ \ \ \ d>-1  \TCItag{90} \\
\left\Vert q_{n}\right\Vert ^{2} &=&\frac{D_{0,1}}{D_{-1,0}}\ \ \ \ \ \ \ \
\ \text{for\ \ \ }d=-1  \TCItag{91}
\end{eqnarray}%
where $D_{d+1,\gamma +1}$ is also given in (B6).

\bigskip

c$%
{{}^\circ}%
)\ $we suppose $-1\leq d<\gamma -1\ \ \Rightarrow \gamma >0$

\qquad

\qquad We are now in the domain where equations (65-66) give from (70-71-72)%
\begin{equation}
\left\{ 
\begin{array}{c}
N_{d}\left( z,y\right) \\ 
K_{d}\left( z,\eta _{k}^{\ast }\right)%
\end{array}%
\right\} \Rightarrow -\sum_{i=d+1}^{m-n+\gamma }\left\{ 
\begin{array}{c}
t_{i}\left( y\right) \\ 
p_{i}^{\ast }\left( \eta _{k}\right)%
\end{array}%
\right\} +\sum_{i=1}^{M_{2}}\beta _{i}\ \left\{ 
\begin{array}{c}
A_{d}\left( x_{i}^{\ast },y\right) \\ 
N_{d}^{\ast }\left( \eta _{k},x_{i}\right)%
\end{array}%
\right\}  \tag{92}
\end{equation}%
This relation leads to biorthogonal polynomials which contains a determinant
of size

$\left( L_{1}+M_{2}+\gamma -d\right) \times \left( L_{2}+M_{1}\right) $%
\begin{eqnarray}
&&q_{n}\left( z;\xi _{i},\eta _{i}^{\ast };y_{i},x_{i}^{\ast }\right) =\frac{%
\dprod\limits_{i=1}^{M_{1}}\left( z-y_{i}\right) }{\dprod%
\limits_{i=1}^{L_{1}}\left( z-\xi _{i}\right) }\ \frac{\left( -\right)
^{\gamma -d-1}\ h_{d}}{D_{d,\gamma }}  \nonumber \\
&&\left\vert 
\begin{array}{cccc}
N_{d}\left( z,y_{i}\right) & t_{\left[ \gamma -1,d+1\right] }\left(
y_{i}\right) & N_{d}\left( \xi _{j},y_{i}\right) & A_{d}\left( x_{l}^{\ast
},y_{i}\right) \\ 
K_{d}\left( z,\eta _{k}^{\ast }\right) & p_{\left[ \gamma -1,d+1\right]
}^{\ast }\left( \eta _{k}\right) & K_{d}\left( \xi _{j},\eta _{k}^{\ast
}\right) & N_{d}^{\ast }\left( \eta _{k},x_{l}\right)%
\end{array}%
\right\vert \ \ \ \ \   \nonumber \\
\text{for\ \ \ \ }\ d &>&-1  \TCItag{93} \\
&&q_{n}\left( z;\xi _{i},\eta _{i}^{\ast };y_{i},x_{i}^{\ast }\right) =\frac{%
\dprod\limits_{i=1}^{M_{1}}\left( z-y_{i}\right) }{\dprod%
\limits_{i=1}^{L_{1}}\left( z-\xi _{i}\right) }\ \frac{\left( -\right)
^{\gamma -1}}{D_{-1,\gamma }}  \nonumber \\
&&\left\vert 
\begin{array}{cccc}
N_{-1}\left( z,y_{i}\right) & t_{\left[ \gamma -1,0\right] }\left(
y_{i}\right) & N_{-1}\left( \xi _{j},y_{i}\right) & A_{-1}\left( x_{l}^{\ast
},y_{i}\right) \\ 
0 & p_{\left[ \gamma -1,0\right] }^{\ast }\left( \eta _{k}\right) & 0 & 
N_{-1}^{\ast }\left( \eta _{k},x_{l}\right)%
\end{array}%
\right\vert \ \ \ \ \   \nonumber \\
\text{for \ }\ d &=&-1  \TCItag{94}
\end{eqnarray}%
where 
\begin{eqnarray}
t_{\left[ \gamma -1,d+1\right] }\left( y_{i}\right) &=&t_{\gamma -1}\left(
y_{i}\right) \ \ t_{\gamma -2}\left( y_{i}\right) ...t_{d+1}\left(
y_{i}\right)  \TCItag{95} \\
p_{\left[ \gamma -1,d+1\right] }^{\ast }\left( \eta _{k}\right) &=&p_{\gamma
-1}^{\ast }\left( \eta _{k}\right) \ \ p_{\gamma -2}^{\ast }\left( \eta
_{k}\right) ...p_{d+1}^{\ast }\left( \eta _{k}\right)  \TCItag{96}
\end{eqnarray}%
and$\ D_{d,\gamma }\ $is given in (B6)\ for $d>-1\ $and$\ D_{-1,\gamma }\ $%
is given in (B7). The pseudonorm is found from (92),\ with $m=n,\ $as%
\begin{eqnarray}
\left\Vert q_{n}\right\Vert ^{2} &=&\left( -\right) ^{\gamma -d}\ h_{d}\ 
\frac{D_{d+1,\gamma +1}}{D_{d,\gamma }}\ \ \ \ \ \ \ \ d>-1  \TCItag{97} \\
\left\Vert q_{n}\right\Vert ^{2} &=&\left( -\right) ^{\gamma }\ \ \frac{%
D_{0,\gamma +1}}{D_{-1,\gamma }}\ \ \ \ \ \ \ \ \ \ \ \ \ \ \ d=-1 
\TCItag{98}
\end{eqnarray}%
where $D_{d+1,\gamma +1}$ is also given in (B6).

\bigskip

\bigskip

d$%
{{}^\circ}%
)\ $we suppose $d<-1,\ \gamma \geq 0$

\bigskip

Equations (92) are still valid but we have to apply (63). The determinant is$%
\ $of size

$\left( L_{1}+M_{2}+\gamma -d\right) \times \left( L_{2}+M_{1}\right) $ 
\begin{eqnarray}
&&q_{n}\left( z;\xi _{i},\eta _{i}^{\ast };y_{i},x_{i}^{\ast }\right) =\frac{%
\dprod\limits_{i=1}^{M_{1}}\left( z-y_{i}\right) }{\dprod%
\limits_{i=1}^{L_{1}}\left( z-\xi _{i}\right) }\ \frac{\left( -\right)
^{\gamma -d}}{D_{d,\gamma }}  \nonumber \\
&&\left\vert 
\begin{array}{ccccc}
N_{-1}\left( z,y_{i}\right) & t_{\left[ \gamma -1,0\right] }\left(
y_{i}\right) & p_{\left[ 0,-d-2\right] }\left( y_{i}\right) & N_{-1}\left(
\xi _{j},y_{i}\right) & A_{-1}\left( x_{l}^{\ast },y_{i}\right) \\ 
0 & p_{\left[ \gamma -1,0\right] }^{\ast }\left( \eta _{k}\right) & 0 & 0 & 
N_{-1}^{\ast }\left( \eta _{k},x_{l}\right)%
\end{array}%
\right\vert  \nonumber \\
\text{for \ \ }\gamma &>&0\ \ \ \ \   \nonumber \\
&&  \TCItag{99} \\
&&q_{n}\left( z;\xi _{i},\eta _{i}^{\ast };y_{i},x_{i}^{\ast }\right) =\frac{%
\dprod\limits_{i=1}^{M_{1}}\left( z-y_{i}\right) }{\dprod%
\limits_{i=1}^{L_{1}}\left( z-\xi _{i}\right) }\ \frac{\left( -\right) ^{-d}%
}{D_{d,0}}  \nonumber \\
&&\left\vert 
\begin{array}{cccc}
N_{-1}\left( z,y_{i}\right) & p_{\left[ 0,-d-2\right] }\left( y_{i}\right) & 
N_{-1}\left( \xi _{j},y_{i}\right) & A_{-1}\left( x_{l}^{\ast },y_{i}\right)
\\ 
0 & 0 & 0 & N_{-1}^{\ast }\left( \eta _{k},x_{l}\right)%
\end{array}%
\right\vert \ \ \ \ \ \ \ \ \   \nonumber \\
\text{for\ }\gamma &=&0  \nonumber \\
&&  \TCItag{100}
\end{eqnarray}%
where%
\begin{equation}
p_{\left[ 0,-d-2\right] }\left( y_{i}\right) =p_{0}\left( y_{i}\right) \ \
p_{_{1}}\left( y_{i}\right) ...p_{-d-2}\left( y_{i}\right)  \tag{101}
\end{equation}%
and where$\ D_{d,\gamma }\ $is given in (B7)\ if $\gamma >0\ $and$\ D_{d,0}$
\ is given in (B8). Finally, the pseudonorm in that case is obtained from
(92), with $m=n,\ $as%
\begin{equation}
\left\Vert q_{n}\right\Vert ^{2}=\left( -\right) ^{\gamma -d-1}\ \ \frac{%
D_{d+1,\gamma +1}}{D_{d,\gamma }}\ \ \ \ \ \ \   \tag{102}
\end{equation}%
where $D_{d+1,\gamma +1}$ is given in (B7).

\bigskip

\bigskip

We note that the pseudonorm can be written, for all values of $d$, and $%
\gamma \ \ $as%
\begin{equation}
\left\Vert q_{n}\right\Vert ^{2}=\left( -\right) ^{\gamma -d}\ \ h_{d,\gamma
}\ \ \frac{D_{d+1,\gamma +1}}{D_{d,\gamma }}  \tag{103}
\end{equation}%
with the convention 
\begin{eqnarray}
h_{d,\gamma } &=&h_{Inf\left( d,\gamma \right) }\ \ \ \ \ \ \ \ \ \ \ \ d\ \ 
\text{and}\ \ \gamma \geq 0  \nonumber \\
h_{d,\gamma } &=&-1\ \ \ \ \ \ \ \ \ \ \ \ \ \ \ \ \ \ \ d\ \ \text{or}\ \
\gamma <0  \nonumber \\
h_{d,\gamma } &=&1\ \ \ \ \ \ \ \ \ \ \ \ \ \ \ \ \ \ \ d\ \ \text{and\ \ }%
\gamma <0  \TCItag{104}
\end{eqnarray}

\bigskip

\bigskip

\section{The integrals $I_{N}$\ and $J_{N}\ \ \ $}

\bigskip

As shown in (14-15), the expression for $I_{N}$ is related to the product of
the pseudonorms from $0$ to $N-1.\ $Since the pseudonorms are essentially
the ratio of two determinants $\frac{D_{d+1,\gamma +1}}{D_{d,\gamma }}$ , we
obtain from (58-59) 
\begin{equation}
I_{N}=\left( -\right) ^{N\left( L_{1}-M_{1}-L_{2}+M_{2}\right) }\ \frac{%
\dprod\limits_{i=0}^{N-1}h_{i+L_{1}-M_{1},i+\ L_{2}-M_{2}}}{%
\dprod\limits_{i=0}^{N-1}h_{i}}\ \ \ \frac{D_{N+L_{1}-M_{1},\ N+L_{2}-M_{2}}%
}{D_{L_{1}-M_{1},L_{2}-M_{2}}}\ \ \ \ \ \ \ \   \tag{105}
\end{equation}%
with the convention (104) and where the determinants\ $%
D_{L_{1}-M_{1},L_{2}-M_{2}}$ and $D_{N+L_{1}-M_{1},\ N+L_{2}-M_{2}}$ are
given in the previous section depending of the values of $N,\ L_{1}-M_{1}$
and $L_{2}-M_{2}.$ In the special case where $L_{1}=M_{1}$ and $L_{2}=M_{2}$
the integrals $I_{N}$ are simply%
\begin{equation}
I_{N}=\frac{D_{N-1}}{D_{-1}}  \tag{106}
\end{equation}%
where%
\begin{eqnarray}
D_{N-1} &=&D_{N,N}=\left\vert 
\begin{array}{cc}
N_{N-1}\left( \xi _{j},y_{i}\right) & A_{N-1}\left( x_{l}^{\ast
},y_{i}\right) \\ 
K_{N-1}\left( \xi _{j},\eta _{k}^{\ast }\right) & N_{N-1}^{\ast }\left( \eta
_{k},x_{l}\right)%
\end{array}%
\right\vert  \TCItag{107} \\
D_{-1} &=&D_{0,0}=\left\vert 
\begin{array}{cc}
N_{-1}\left( \xi _{j},y_{i}\right) & A_{-1}\left( x_{l}^{\ast },y_{i}\right)
\\ 
0 & N_{-1}^{\ast }\left( \eta _{k},x_{l}\right)%
\end{array}%
\right\vert  \nonumber \\
&=&N_{-1}\left( \xi _{j},y_{i}\right) \ \ N_{-1}^{\ast }\left( \eta
_{k},x_{l}\right)  \TCItag{108}
\end{eqnarray}%
with%
\begin{equation}
N_{-1}\left( \xi _{j},y_{i}\right) =\left\vert 
\begin{array}{ccc}
\frac{1}{y_{1}-\xi _{1}} & ... & \frac{1}{y_{1}-\xi _{M_{1}}} \\ 
... & ... & ... \\ 
\frac{1}{y_{M_{1}}-\xi _{1}} & ... & \frac{1}{y_{M_{1}}-\xi _{M_{1}}}%
\end{array}%
\right\vert =\left( -\right) ^{\frac{M_{1}\left( M_{1}-1\right) }{2}}\frac{%
\Delta \left( y\right) \ \Delta \left( \xi \right) }{\dprod\limits_{i,j}%
\left( y_{i}-\xi _{j}\right) }  \tag{109}
\end{equation}%
The result (106-107-108-109)\ is nothing but equation (18).

\bigskip

\bigskip

In order to obtain the resolvent $J_{N}$ from $I_{N},$ we must perform the\
derivatives%
\begin{equation}
\left[ \dprod\limits_{i=1}^{M_{2}}\left( -\frac{\partial }{\partial \eta
_{i}^{\ast }}\right) \ \dprod\limits_{i=1}^{M_{1}}\left( -\frac{\partial }{%
\partial \xi _{i}}\right) \ \ \ \frac{\dprod\limits_{i,j}^{M_{2}}\left(
x_{i}^{\ast }-\eta _{j}^{\ast }\right) \ \dprod\limits_{i,j}^{M_{1}}\left(
y_{i}-\xi _{j}\right) }{\Delta \left( x^{\ast }\right) \ \Delta \left(
y\right) \ \Delta \left( \eta ^{\ast }\right) \ \Delta \left( \xi \right) }\
\ D_{n}\right] _{x_{i}^{\ast }=\eta _{i}^{\ast },\ y_{i}=\xi _{i}}  \tag{110}
\end{equation}%
This operation can be decomposed into the derivatives in $\xi _{i}\ $%
involving\ the $M_{1}\ $first column of $D_{n}$ while the derivatives in $%
\eta _{i}^{\ast }$ involve the $M_{2}$ last rows of $D_{n}.$\ Both
operations are similar and we describe here the derivatives in $\xi _{i}.\ $%
We define%
\begin{equation}
R\left( y_{k},\xi _{j}\right) =\frac{\dprod\limits_{k\neq j}\left( y_{k}-\xi
_{j}\right) }{\Delta \left( y\right) \ \Delta \left( \xi \right) }  \tag{111}
\end{equation}%
while the factors $\left( y_{i}-\xi _{i}\right) \ $are distributed over the
columns of $D_{n}.$ Given $I\subset $ we write%
\begin{equation}
\sum_{I}\left[ \dprod\limits_{i\notin I}\left( -\frac{\partial }{\partial
\xi _{i}}\right) R\left( y_{k},\xi _{j}\right) \right] _{y_{i}=\xi _{i}}\ %
\left[ \dprod\limits_{i\in I}\left( -\frac{\partial }{\partial \xi _{i}}%
\right) \left\{ \dprod\limits_{k=1}^{M_{1}}\left( y_{k}-\xi _{k}\right) \
D_{n}\ \right\} \right] _{y_{i}=\xi _{i}}  \tag{112}
\end{equation}%
The right $\left[ .\right] $ is easily calculated since for the columns $%
i\notin I$ we obtain $1$ on the diagonal and $0$ otherwise, and for the
columns $i\in I\ $we obtain $H_{n}\left( \xi _{i},\xi _{i}\right) $ on the
diagonal, the remainder being unchanged except for $y_{k}=\xi _{k}.\ $We
call $D_{n}\left( I\right) $\ the corresponding subdeterminant with the
indices $i\in I.$ Now, we calculate the left $\left[ .\right] ;$ for any $%
J\subset \left( \xi _{1},...,\xi _{M_{1}}\right) $ we have%
\begin{equation}
\left[ \dprod\limits_{i\notin I}\left( -\frac{\partial }{\partial \xi _{i}}%
\right) R\left( y_{k},\xi _{j}\right) \right] _{y_{i}=\xi _{i}}=0\ \ \ \ \ 
\text{if card}\left( J\right) \ \text{is odd}  \tag{113}
\end{equation}%
and if card$\left( J\right) \ $is even%
\begin{equation}
\left[ \dprod\limits_{i\notin I}\left( -\frac{\partial }{\partial \xi _{i}}%
\right) R\left( y_{k},\xi _{j}\right) \right] _{y_{i}=\xi _{i}}=\left(
-\right) ^{\frac{M_{1}\left( M_{1}-1\right) }{2}\ }\sum_{all\ pairings\ in\
J}\dprod\limits_{\left( j,k\right) }\left( \frac{-1}{\left( \xi _{k}-\xi
_{j}\right) ^{2}}\right)  \tag{114}
\end{equation}%
Altogether, we have obtained%
\begin{eqnarray}
&&\left( -\right) ^{\frac{M_{1}\left( M_{1}-1\right) }{2}\ }\sum_{I\ even}%
\left[ \sum_{all\ pairings\ in\ J}\dprod\limits_{\left( i,j\right) }\left( 
\frac{-1}{\left( \xi _{i}-\xi _{j}\right) ^{2}}\right) \right] \ D_{n}\left(
I\right)  \nonumber \\
&=&\left( -\right) ^{\frac{M_{1}\left( M_{1}-1\right) }{2}\ }"D_{n}" 
\TCItag{115}
\end{eqnarray}%
where $"D_{n}"$ is the determinant obtained from $D_{n}$ after the following
changes: the diagonal is $H_{n}\left( \xi _{i},\xi _{i}\right) ,$ all
variables $y_{i}\ $are changed into $\xi _{i},$ finally " " means that all
double poles at $\xi _{k}=\xi _{j}$ are ignored as we develop the
determinant. In fact, there is no single poles at $\xi _{k}=\xi _{j}$
either, since the residues are zero.

\bigskip The same result applies for the derivatives $\dprod%
\limits_{i=1}^{M_{2}}\left( -\frac{\partial }{\partial \eta _{i}^{\ast }}%
\right) _{x_{i}^{\ast }=\eta _{i}^{\ast }}$ and this gives $J_{N}$ as in
(20).

\bigskip

\bigskip

\bigskip

\section{Appendix A}

\bigskip

We compute the large $y$ behaviour of the function\ $t_{n}\left( y\right) \ $%
and of the kernels $N_{n}\left( \xi ,y\right) \ $and\ $A_{n}\left( x^{\ast
},y\right) \ $defined in (27)\ and in (31).\ The formal power series for $%
t_{n}\left( y\right) $ when $y\rightarrow \infty $ can be written as%
\begin{eqnarray}
t_{n}\left( y\right) &=&-\sum_{p=0}^{\infty }\ a_{p}\ \frac{1}{y^{p+1}} 
\TCItag{A1} \\
a_{p} &=&\int d\mu \left( z,z^{\ast }\right) \ p_{n}^{\ast }\left( z\right)
\ z^{p}  \TCItag{A2}
\end{eqnarray}%
Clearly, $a_{p}=0\ $for $p<n.\ $Let us write%
\begin{equation}
z^{p}=\sum_{k=0}^{p}\alpha _{pk}\ p_{k}\left( z\right) ,\ \ \ \ \alpha
_{pk}\ =0\ \text{if\ }k>p,\ \ \ \ \alpha _{pp}\ =1  \tag{A3}
\end{equation}%
Consequently,%
\begin{eqnarray}
t_{n}\left( y\right) &=&-h_{n}\sum_{p=n}^{\infty }\ \alpha _{pn}\ \frac{1}{%
y^{p+1}}  \TCItag{A4} \\
t_{n}\left( y\right) &\sim &-\frac{h_{n}}{y^{n+1}}\ \text{as}\ y\rightarrow
\infty  \TCItag{A5}
\end{eqnarray}%
Now, if we insert the expansion (A4)\ for $t_{n}\left( y\right) $ and if we
use (A3), we obtain%
\begin{equation}
\sum_{n=0}^{\infty }\frac{p_{n}\left( \xi \right) \ t_{n}\left( y\right) }{%
h_{n}}=\frac{1}{\xi -y}  \tag{A6}
\end{equation}%
If this expression is reported in the definition of $N_{n}\left( \xi
,y\right) $ we obtain 
\begin{eqnarray}
N_{n}\left( \xi ,y\right) &=&-\sum_{p=n+1}^{\infty }\frac{p_{p}\left( \xi
\right) \ t_{p}\left( y\right) }{h_{p}}  \TCItag{A7} \\
N_{n}\left( \xi ,y\right) &\sim &\frac{p_{n+1}\left( \xi \right) }{y^{n+2}}\
\ \ \text{as}\ y\rightarrow \infty  \TCItag{A8}
\end{eqnarray}%
Similarly, we find for $Q\left( x^{\ast },y\right) $ defined in (30)%
\begin{equation}
Q\left( x^{\ast },y\right) =\sum_{n=0}^{\infty }\frac{t_{n}^{\ast }\left(
x\right) \ t_{n}\left( y\right) }{h_{n}}  \tag{A9}
\end{equation}%
so that%
\begin{eqnarray}
A_{n}\left( x^{\ast },y\right) &=&-\sum_{p=n+1}^{\infty }\frac{t_{p}^{\ast
}\left( x\right) \ t_{p}\left( y\right) }{h_{p}}  \TCItag{A10} \\
A_{n}\left( x^{\ast },y\right) &\sim &\frac{t_{n+1}^{\ast }\left( x\right) }{%
y^{n+2}}\ \ \ \text{as}\ y\rightarrow \infty  \TCItag{A11}
\end{eqnarray}

\bigskip

\bigskip

\bigskip

\section{Appendix B}

\bigskip

We collect the various forms of the determinant $D_{d,\gamma }$ according to
the values of $d$ and $\gamma .$

\bigskip

\subsection{$\mathbf{\protect\gamma <0}$}

\textbf{\bigskip }

1a$%
{{}^\circ}%
)$\ \ $d>0$%
\begin{equation}
D_{d,\gamma }=\left\vert 
\begin{array}{cc}
p_{d-1}\left( \xi _{j}\right) & t_{d-1}^{\ast }\left( x_{l}\right) \\ 
... & ... \\ 
p_{0}\left( \xi _{j}\right) & t_{0}^{\ast }\left( x_{l}\right) \\ 
0 & p_{0}^{\ast }\left( x_{l}\right) \\ 
... & ... \\ 
0 & p_{-\gamma -1}^{\ast }\left( x_{l}\right) \\ 
N_{d-1}\left( \xi _{j},y_{i}\right) & A_{d-1}\left( x_{l}^{\ast
},y_{i}\right) \\ 
K_{d-1}\left( \xi _{j},\eta _{k}^{\ast }\right) & N_{d-1}^{\ast }\left( \eta
_{k},x_{l}\right)%
\end{array}%
\right\vert  \tag{B1}
\end{equation}

1b$%
{{}^\circ}%
)\ d=0$%
\begin{equation}
D_{0,\gamma }=\left\vert 
\begin{array}{cc}
0 & p_{0}^{\ast }\left( x_{l}\right) \\ 
... & ... \\ 
0 & p_{-\gamma -1}^{\ast }\left( x_{l}\right) \\ 
N_{-1}\left( \xi _{j},y_{i}\right) & A_{-1}\left( x_{l}^{\ast },y_{i}\right)
\\ 
0 & N_{-1}^{\ast }\left( \eta _{k},x_{l}\right)%
\end{array}%
\right\vert  \tag{B2}
\end{equation}

1c$%
{{}^\circ}%
)\ d<0$%
\begin{equation}
D_{d,\gamma }=\left\vert 
\begin{array}{ccccc}
0 & ... & 0 & 0 & p_{0}^{\ast }\left( x_{l}\right) \\ 
... & ... & ... & ... & ... \\ 
0 & ... & 0 & 0 & p_{-\gamma -1}^{\ast }\left( x_{l}\right) \\ 
p_{0}\left( y_{i}\right) & ... & p_{-d-1}\left( y_{i}\right) & N_{-1}\left(
\xi _{j},y_{i}\right) & A_{-1}\left( x_{l}^{\ast },y_{i}\right) \\ 
0 & ... & 0 & 0 & N_{-1}^{\ast }\left( \eta _{k},x_{l}\right)%
\end{array}%
\right\vert  \tag{B3}
\end{equation}

\bigskip

\subsection{$\mathbf{\protect\gamma \geq 0}$}

\bigskip

\qquad 2a$%
{{}^\circ}%
)\ \ d>\gamma $%
\begin{equation}
D_{d,\gamma }=\left\vert 
\begin{array}{cc}
p_{d-1}\left( \xi _{j}\right) & t_{d-1}^{\ast }\left( x_{l}\right) \\ 
... & ... \\ 
p_{\gamma }\left( \xi _{j}\right) & t_{\gamma }^{\ast }\left( x_{l}\right)
\\ 
N_{d-1}\left( \xi _{j},y_{i}\right) & A_{d-1}\left( x_{l}^{\ast
},y_{i}\right) \\ 
K_{d-1}\left( \xi _{j},\eta _{k}^{\ast }\right) & N_{d-1}^{\ast }\left( \eta
_{k},x_{l}\right)%
\end{array}%
\right\vert \ \   \tag{B4}
\end{equation}

2b$%
{{}^\circ}%
)\ d=\gamma $%
\begin{equation}
D_{d,\gamma }=\left\vert 
\begin{array}{cc}
N_{d-1}\left( \xi _{j},y_{i}\right) & A_{d-1}\left( x_{l}^{\ast
},y_{i}\right) \\ 
K_{d-1}\left( \xi _{j},\eta _{k}^{\ast }\right) & N_{d-1}^{\ast }\left( \eta
_{k},x_{l}\right)%
\end{array}%
\right\vert  \tag{B5}
\end{equation}

2c$%
{{}^\circ}%
)\ \ 0\leq d<\gamma $%
\begin{equation}
D_{d,\gamma }=\left\vert 
\begin{array}{ccc}
t_{\left[ \gamma -1,d\right] }\left( y_{i}\right) & N_{d}\left( \xi
_{j},y_{i}\right) & A_{d}\left( x_{l}^{\ast },y_{i}\right) \\ 
p_{\left[ \gamma -1,d\right] }^{\ast }\left( \eta _{k}\right) & K_{d}\left(
\xi _{j},\eta _{k}^{\ast }\right) & N_{d}^{\ast }\left( \eta
_{k},x_{l}\right)%
\end{array}%
\right\vert  \tag{B6}
\end{equation}%
where $t_{\left[ d,\gamma -1\right] }\left( y_{i}\right) $ and $p_{\left[
d,\gamma -1\right] }^{\ast }\left( \eta _{k}\right) $ are defined in (95-96).

\bigskip

2d$%
{{}^\circ}%
)\ \ d\leq -1,\ \gamma \geq 1$%
\begin{equation}
D_{d,\gamma }=\left\vert 
\begin{array}{cccc}
t_{\left[ \gamma -1,0\right] }\left( y_{i}\right) & p_{\left[ 0,-d-1\right]
}\left( y_{i}\right) & N_{-1}\left( \xi _{j},y_{i}\right) & A_{-1}\left(
x_{l}^{\ast },y_{i}\right) \\ 
p_{\left[ \gamma -1,0\right] }^{\ast }\left( \eta _{k}\right) & 0 & 0 & 
N_{-1}^{\ast }\left( \eta _{k},x_{l}\right)%
\end{array}%
\right\vert  \tag{B7}
\end{equation}%
where $p_{\left[ -d-1,0\right] }\left( y_{i}\right) $ is defined in (101).

\bigskip 2e$%
{{}^\circ}%
)\ \ d\leq -1,\ \gamma =0$%
\begin{equation}
D_{d,\gamma }=\left\vert 
\begin{array}{ccc}
p_{\left[ 0,-d-1\right] }\left( y_{i}\right) & N_{-1}\left( \xi
_{j},y_{i}\right) & A_{-1}\left( x_{l}^{\ast },y_{i}\right) \\ 
0 & 0 & N_{-1}^{\ast }\left( \eta _{k},x_{l}\right)%
\end{array}%
\right\vert  \tag{B8}
\end{equation}

\bigskip

\section{\protect\bigskip References}

1.\ \ \ \ Dijkgraaf R.and Vafa C.: Nucl. Phys. \textbf{B644},\ 3-39 (2002)

\noindent 2.\ \ \ \ Verbaarschot J.J.M.\ and Wettig T.: Ann. Rev. Nucl.
Part. Sci. \textbf{50},\newline
\hspace*{1cm}343-410 (2000)

\noindent 3$.$\ \ \ \ Akemann G.: Phys. Lett. \textbf{B547}, 100-108 (2002)

\noindent 4.\ \ \ \ Szeg\"{o} G.: orthogonal polynomials, Am. Math. Soc.
Colloquium publica\newline
\hspace*{1cm}tions \textbf{23},\ Providence (1975)

\noindent 5$.$\ \ \ \ Brezin E. and Hikami S.: Commun. Math. Phys. \textbf{%
214}, 111-135 (2000)

\noindent 6$.\ \ \ \ $Uvarov\ V.B.: USSR Comput. Math. Phys. \textbf{9}, n$%
{{}^\circ}%
)$ 6, 25-36\ (1969)

\noindent 7$.$\ \ \ \ Fyodorov Y. and Strahov E.: J. of Phys.\ \textbf{A36},
3203-3213 (2003)

\noindent 8$.$\ \ \ \ Baik J., Deift P. and Strahov E.: J. Math. Phys. 
\textbf{44}, 3657-3670 (2003)

\noindent 9.\ \ \ \ Akemann G. and Vernizzi G.: Nucl. Phys. \textbf{B660},
532-556 (2003)

\noindent 10.$\ \ $Bergere M.C.:\ "orthogonal polynomials for potentials of
two variables with\newline
\hspace*{1cm}external sources", hep-th/0311227

\noindent 11.\ \ Eynard B. and Mehta M.L.: Journ. of Phys. \textbf{A31},
4449-4456 (1998)

\noindent 12.$\ \ $Bergere M.C.: 4th Eurogrid meeting, les Houches (march
2004)

\noindent 13.\ \ Berenstein D., Maldacena J. and Nastase H.: JHEP \textbf{%
0204}, 013 (2002

\noindent 14. \ Eynard B. and Kristjansen C.: JHEP \textbf{0210}, 027 (2002)

\noindent 15. \ Beisert N., Kristjansen C., Plefka J., Semenoff G.W. and%
\newline
\hspace*{1cm}Staudacher M.: Nucl.\ Phys. \textbf{B650}, 125-161\ \ (2003)

\end{document}